\documentclass[journal]{IEEEtran}

\usepackage{graphicx}
\usepackage{epstopdf}
\usepackage[fleqn]{amsmath}
\usepackage{amsthm}
\usepackage{algorithm}
\usepackage{algorithmic}
\usepackage{longtable}
\usepackage{array}
\usepackage{setspace}
\usepackage{bm}
\usepackage{subfigure}
\usepackage{color}

\ifCLASSINFOpdf

\else

\fi

\begin{document}

\renewcommand{\baselinestretch}{1.00}

\title{\huge Hyper-parameter Optimization for Wireless Network Traffic Prediction Models with A Novel Meta-Learning Framework}

\vspace{-2.8cm}

\author{Liangzhi Wang, 
~Jie~Zhang, \emph{Senior Member, IEEE},
~Yuan~Gao,
~Jiliang~Zhang, \emph{Senior Member, IEEE},
~Guiyi~Wei, \emph{Member, IEEE},
~Haibo~Zhou, \emph{Senior Member, IEEE},
~Bin~Zhuge,
~and~Zitian~Zhang

\thanks{
Liangzhi Wang is with the Department of Electronic and Electrical Engineering, University of Sheffield, Sheffield, S10 2TN, UK.

%(e-mail: jie.zhang@sheffield.ac.uk)
Jie Zhang is with the Department of Electronic and Electrical Engineering, University of Sheffield, Sheffield, S10 2TN, UK, and also with Ranplan Wireless Network Design Ltd., Cambridge, CB23 3UY, UK.

%(e-mail:gaoyuansie@shu.edu.cn; jinyanliang@staff.shu.edu.cn)
Yuan Gao is with the School of Communication and Information Engineering, Shanghai University, Shanghai, China.

%(e-mail: zhangjiliang1@mail.neu.edu.cn)
Jiliang Zhang is with College of Information Science and Engineering, Northeastern University, Shenyang, China.

Guiyi Wei, Bin Zhuge, and Zitian Zhang are with School of Information and Electronic Engineering, Zhejiang Gongshang University, Hangzhou, China.

Haibo Zhou is with the Department of Electrical and Computer Engineering, Nanjing University, Nanjing, China.

Zitian Zhang is the corresponding author. 

(e-mail: zitian.zhang@mail.zjgsu.edu.cn) 
%Copyright (c) 20xx IEEE. Personal use of this material is permitted. However, permission to use this material for any other purposes must be obtained from the IEEE by sending a request to pubs-permissions@ieee.org.
}

}

\markboth{IEEE submitted manuscript}%
{Submitted paper}

\maketitle

\renewcommand{\baselinestretch}{1.00}

\begin{abstract}
This paper proposes a novel meta-learning based hyper-parameter optimization framework for wireless network traffic prediction (NTP) models. 
The primary objective is to accumulate and leverage the acquired hyper-parameter optimization experience, enabling the rapid determination of optimal hyper-parameters for new tasks. 
In this paper, an attention-based deep neural network (ADNN) is employed as the base-learner to address specific NTP tasks. The meta-learner is an innovative framework that integrates meta-learning with the k-nearest neighbor algorithm (KNN), genetic algorithm (GA), and gated residual network (GRN). 
Specifically, KNN is utilized to identify a set of candidate hyper-parameter selection strategies for a new task, which then serves as the initial population for GA, while a GRN-based chromosome screening module accelerates the validation of offspring chromosomes, ultimately determining the optimal hyper-parameters. 
Experimental results demonstrate that, compared to traditional methods such as Bayesian optimization (BO), GA, and particle swarm optimization (PSO), the proposed framework determines optimal hyper-parameters more rapidly, significantly reduces optimization time, and enhances the performance of the base-learner. It achieves an optimal balance between optimization efficiency and prediction accuracy.

\end{abstract}


\IEEEpeerreviewmaketitle
\vspace{-2mm}
\section{Introduction}
\IEEEPARstart
{I}{n} recent years, the exponential growth in wireless network traffic demand has imposed significant challenges on network operational efficiency and expenditure \cite{Agiwal2016}.
As a pivotal solution to this challenge, wireless network traffic prediction (NTP) aims to accurately forecast traffic demands across various cities, mobile cells, and even Internet of Things (IoT) terminal devices \cite{Zhang_WCM_2021}. This capability facilitates the efficient management and allocation of network resources, both in real-time and proactively \cite{I. Lohrasbinasab}.
For instance, if the future traffic loads of multiple cooperative access points can be accurately predicted, an optimal sleeping scheme can be implemented to reduce their energy consumption \cite{Jinming Hu2014}. In addition, many emerging intelligent IoT architectures depend on precise device-level traffic prediction to enhance their quality of service (QoS) and quality of experience (QoE) performance \cite{Jiliang}.

Optimizing hyper-parameters in wireless NTP models is crucial, especially for mainstream deep learning-based models. 
Unlike mathematical model-based methods [6]-[14] that use predefined functions to model traffic load patterns and shallow learning-based methods [15]-[20] that construct prediction models using traditional machine learning algorithms, deep learning-based traffic prediction methods [21]-[31] have the potential to reveal complex nonlinear relationships in wireless network traffic loads. This advantage stems from the ability of deep neural networks to automatically extract features. As a result, these data-driven deep learning-based models have achieved state-of-the-art accuracy and have become the most widely used methods in recent years. 
However, their performance is heavily dependent on hyper-parameter quality, including the number of layers, neurons per layer, and learning rate, among others.
In other words, hyper-parameter selection strategy directly impacts how well deep learning models realize their prediction potential.

Therefore, it is necessary to design a tailored hyper-parameter optimization method for NTP field. 
In 5G and beyond mobile networks, the ultra-dense deployment of mobile cells, access points, and IoT devices [1-2] will introduce tens of thousands of wireless NTP tasks, which in turn necessitate tens of thousands of deep learning-based prediction models.
Given the massive number of NTP tasks, determining optimal hyper-parameter selection strategies for each model individually to cope with complex wireless NTP tasks demands significant computational resources and optimization time. This process inevitably escalates the network's operating expenses (OPEX) to an unsustainable level.

However, research on hyper-parameter optimization for wireless NTP models remains underdeveloped. 
On the one hand, existing studies on NTP primarily focus on proposing algorithms for single-task scenarios to improve prediction accuracy. Only a limited number of studies have explored the impact of hyper-parameter selection on NTP models, and these efforts remain confined to basic comparisons involving a small set of hyper-parameters. A systematic investigation of hyper-parameter optimization method for NTP tasks is still lacking. 
On the other hand, contemporary hyper-parameter optimization methods, such as evolutionary algorithms and Bayesian optimization (BO), effectively reduce computational costs for individual tasks, but remain limited in their ability to generalize beyond single-task scenarios. Specifically, they lack the ability to transfer acquired knowledge across homogeneous problem domains and do not incorporate a meta-optimization mechanism to progressively refine search strategies. This limitation hinders the development of cumulative optimization intelligence, thereby limiting efficiency improvements.

Compared to the existing works, this paper not only explores the predictive potential of a specific NTP model within a given task but, more importantly, systematically investigates hyper-parameter optimization methods for large-scale deep learning-based wireless NTP models. Specifically, it leverages meta-learning principles and integrates the k-nearest neighbor algorithm (KNN), genetic algorithm (GA), and gated residual network (GRN) to develop an innovative and efficient hyper-parameter optimization framework. The main contributions of this study are summarized as follows:

\begin{itemize}
	\item This paper investigates the correlation between wireless NTP tasks and their optimal hyper-parameters, which serves as the foundation for determining the meta-features of meta-samples within the proposed framework. Specifically, by analyzing the information entropy of various hyper-parameters and the conditional entropy of hyper-parameters under different intrinsic characteristics of prediction tasks, meta-features are determined based on the extent to which these characteristics influence optimal hyper-parameters.
	\item This paper incorporates meta-learning into the proposed framework, enabling the framework to transfer hyper-parameter optimization experience across homogeneous problem domains and progressively refine the search strategy. Each wireless NTP task is designated as a base-task, where the state-of-the-art attention-based deep neural network (ADNN) serves as the base-learner. 
The optimization process for determining the most effective hyper-parameters of the base-learner is systematically formulated as a corresponding meta-task.
	\item For large-scale wireless NTP tasks, this paper proposes a novel meta-learning framework that integrates GA and KNN to advance conventional approaches. This integration achieves dual enhancement in both optimization effectiveness and computational efficiency.
Specifically, KNN identifies high-performance candidate hyper-parameter combinations for a new base-learner, providing high-quality initialization for the GA population. Subsequently, GA performs an iterative search, effectively enhancing optimization efficiency and quality.
	\item 
This paper proposes an innovative chromosome screening module based on GRN, which efficiently evaluates GA-generated offspring hyper-parameter selection strategies by leveraging the meta-knowledge from meta-tasks and the meta-features of the corresponding base-task.  
By rapidly filtering these offspring chromosomes, the module significantly reduces the computational overhead associated with individually verifying the fitness of offspring chromosomes. This module alleviates the extensive training burden of new base-learners and conserves substantial computational resources as well as optimization time, thereby greatly enhancing the efficiency of hyper-parameter optimization.
	\item This framework is designed for large-scale wireless NTP tasks, effectively capturing the complexities of numerous small cells and data patterns. 
Extensive experiments demonstrate that the proposed framework rapidly determines optimal hyper-parameters, significantly reducing optimization time. In particular, it effectively balances optimization efficiency and prediction accuracy, achieving state-of-the-art performance. Furthermore, the framework exhibits strong robustness across various base-learners.
\end{itemize}

The rest of this paper is organized as follows. 
Existing work in the field of wireless NTP is reviewed in Section II.
Section III introduces the real-world wireless network traffic records adopted in this paper, followed by our statistical analysis regarding the optimal hyper-parameter selection strategies of the prediction models. 
Section IV describes the proposed meta-learning based hyper-parameter optimization framework for wireless NTP in detail. 
Section V presents the experimental setup and evaluates the performance of the proposed framework. 
Finally, Section VI concludes the paper.

%\vspace{-2mm}
\section{Related works}
Existing wireless NTP methods can be broadly classified into three main categories: mathematical model-based methods, shallow learning-based methods, and deep learning-based methods.

Mathematical model-based methods align historical wireless network traffic data with specific mathematical or statistical models, deriving predictions based on the models' evolving trends or probabilistic distributions.
The authors in \cite{Liangzhi} proposed a user-behavior-based NTP method that enhances the interpretability of the NTP model by capturing user behavior across major daily periods.
In \cite{Ge Xiaohu 2004}, the authors focused on the self-similar network traffic and leveraged the theory of $\alpha$-stable processes to extract the traffic patterns in wireless networks.
The linear auto-regressive integrated moving average (ARIMA) based models were employed in \cite{Yantai Shu 2005} and \cite{Fengli Xu 2016} to capture the seasonal correlations in network traffic.
To further improve the ARIMA model's performance, an ARIMA model aided by entropy theory was proposed in \cite{Rongpeng Li 2014}. 
In addition to these models, the ON-OFF model \cite{yuan13}, the Kalman function \cite{Kalman function P2 right}, the covariance function \cite{yuan14}, and the Holt-Winter's exponential smoothing model \cite{yuan12} were also used to fit the temporal and/or spatial characteristics of wireless traffic loads. 
Mathematical model-based approaches offer advantages such as simplicity, low computational complexity, and no complex hyper-parameters. However, its limited learning capacity, and the highly intricate nature of cell-level wireless network traffic make them less suitable for accurate cell-level wireless NTP tasks.

For the second category, numerous shallow learning algorithms, like linear regression \cite{yuan15}, support vector regression \cite{yuan18}, compressive sensing \cite{yuan16} \cite{yuan17}, Gaussian processes \cite{yuan21}, and principal component analyses \cite{yuan19} have been proposed for NTP tasks. 
These algorithms typically involve relatively few hyper-parameters.
However, their performance and hyper-parameter selection heavily depend on expert knowledge of wireless NTP, which may limit their applicability. Consequently, in the absence of such expertise, these models may perform suboptimally. Moreover, their limited learning capacity is often insufficient to capture the complex characteristics of wireless network traffic, potentially resulting in low prediction accuracy even with abundant training data.

In recent years, deep learning techniques have been increasingly applied to wireless NTP field.
In \cite{yuan22}, the authors proposed a deep belief network and Gaussian models (DBNG) to predict the cell-level wireless network traffic loads. In \cite{yuan25}, C. Qiu \textit {et al.} proposed a multi-task learning approach based on the long short-term memory (LSTM) network, which improves the model's prediction accuracy by exploring similarities and differences of traffic patterns among neighboring mobile cells. 
With reduced connection probability between neurons, a random connectivity LSTM network based traffic prediction model was proposed in \cite{yuan26} to decrease the model's training complexity. 
A multi-scale deep echo-state network (ESN)-based prediction model was proposed in \cite{ESN} to learn the trends and characteristics of network traffic at different temporal scales. 
Regarding historical traffic loads generated in multiple base stations as input, the convolutional neural network (CNN) and convolutional LSTM-based NTP models were proposed in \cite{CNN} and \cite{Convolutional LSTM}, respectively. 
In \cite{Bi LSTM}, the authors proposed to decompose the network traffic loads into several product function components using the local mean decomposition method, each of which is then predicted with a bidirectional LSTM network model. 
Y. Hu \textit {et al.} \cite{Hu et al. Transformer P3 left} proposed to utilize attention mechanism to depict the spatial-temporal characteristics of wireless traffic patterns and thus presented a transformer based prediction model. 
Z. Wang \textit {et al.} \cite{GNN} integrated the network spatial information with the cell-level wireless network traffic load series and proposed a graph neural network (GNN)-based NTP model. By combining attention and convolution mechanisms into traffic analysis, a multi-view spatial-temporal graph network (MVSTGN)-based prediction model was proposed in \cite{MVSTGN} to learn diverse global spatial-temporal dependencies of cellular traffic loads. 
These studies have demonstrated the capability of deep learning to uncover hidden features in wireless traffic patterns, achieving state-of-the-art prediction performance.  
However, most of them focused on designing sophisticated prediction models for various NTP tasks while neglecting the problem of optimal hyper-parameter selection. 
Specifically, they overlook how to determine the optimal hyper-parameters based on the intrinsic characteristics of the prediction task or leverage knowledge gained from other tasks.
In \cite{Fuyou}, authors made some initial attempts to elevate a new cell-level traffic prediction model's performance by providing it proper initial weight vector with the help of initial weight vector selection strategies of previous prediction models. Nevertheless, optimizing the prediction models' hyper-parameters has also not been addressed in \cite{Fuyou}.

%\vspace{-2mm}
\section{Dataset and preliminary analyses}

\subsection{Wireless network traffic records}
In this paper, wireless network traffic data in the "Telecom Italia Bia Data Challenge" from 01/11/2013 to 01/01/2014 in Milan \cite{Milan Data} serves as the dataset.
In spatial dimension, Milan city is covered by 10000 grids, each of which possesses a size of $235m\times235m$.
Each traffic record in the dataset includes information on the occurrence time, network traffic volume, and grid ID. Since the coverage area of an urban base station is approximately the same size as a grid, each grid is referred to as a mobile cell and represented by the corresponding grid ID \cite{Milan Data}.
The time span of the dataset is divided into 8928 intervals with a duration of ten minutes. Traffic load of the $p$-th cell ($p=1,\;...,\;10000$) during the $t$-th time interval ($t=1,\;...,\;8928$) can be acquired as $L_p[t]$ while the traffic load series of mobile cell $p$ can be denoted as ${{\bf{L}}_p} = ({L_p}[1],\;{L_p}[2],\;...,\;{L_p}[8928])$. In order to analyze characteristics of traffic load series generated in different cells with a uniform scale, we normalized elements in ${{\bf{L}}_p}$ into the range of [0,1] using the max-min normalization method as follows.
\begin{small}
	\begin{equation}
	{\widetilde L_p}[t] = \frac{{{L_p}[t] - \min ({{\bf{L}}_p})}}{{\max ({{\bf{L}}_p}) - \min ({{\bf{L}}_p})}},
	\label{eq:eq101}
	\end{equation}
\end{small}
%\vspace{-0.2cm}
\noindent
where $\max ({{\bf{L}}_p})$ and $\min ({{\bf{L}}_p})$ are the largest and smallest elements in ${{\bf{L}}_p}$, respectively. Accordingly, the normalized traffic load series of mobile cell $p$ is denoted as ${\widetilde {\bf{L}}_p}$.

\begin{figure*}
\vspace{-0.5cm} 
	\centering
	\subfigure[]{
		\begin{minipage}[t]{0.31\linewidth}
			\centering
			\includegraphics[width=2.4in]{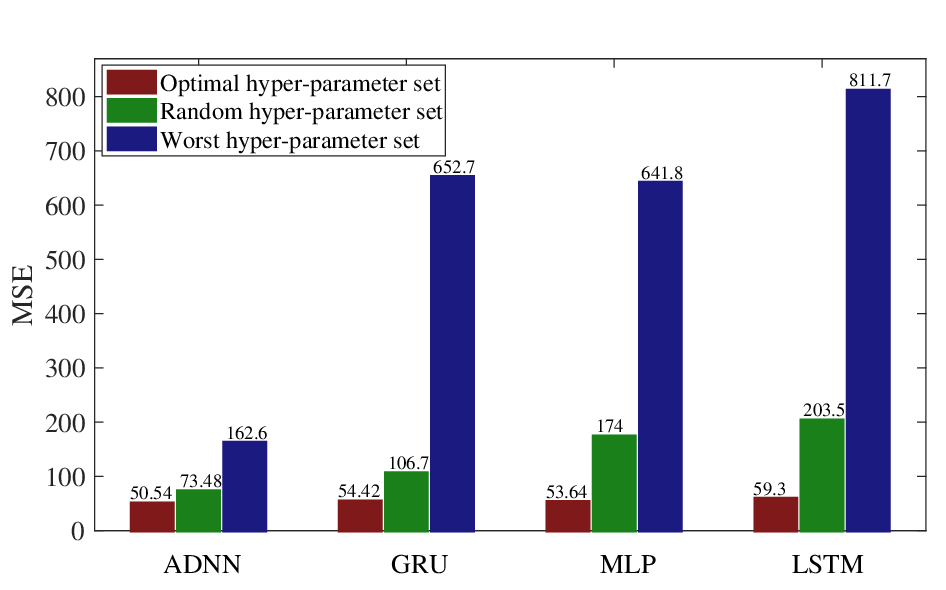}
		\end{minipage}
	}
	\subfigure[]{
		\begin{minipage}[t]{0.31\linewidth}
			\centering
			\includegraphics[width=2.4in]{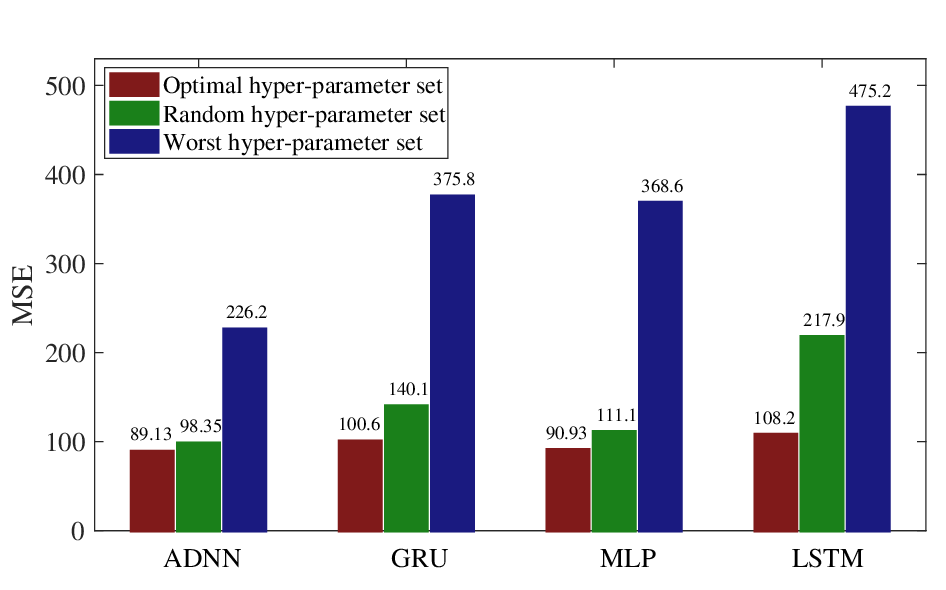}
		\end{minipage}
	}
	\subfigure[]{
		\begin{minipage}[t]{0.31\linewidth}
			\centering
			\includegraphics[width=2.4in]{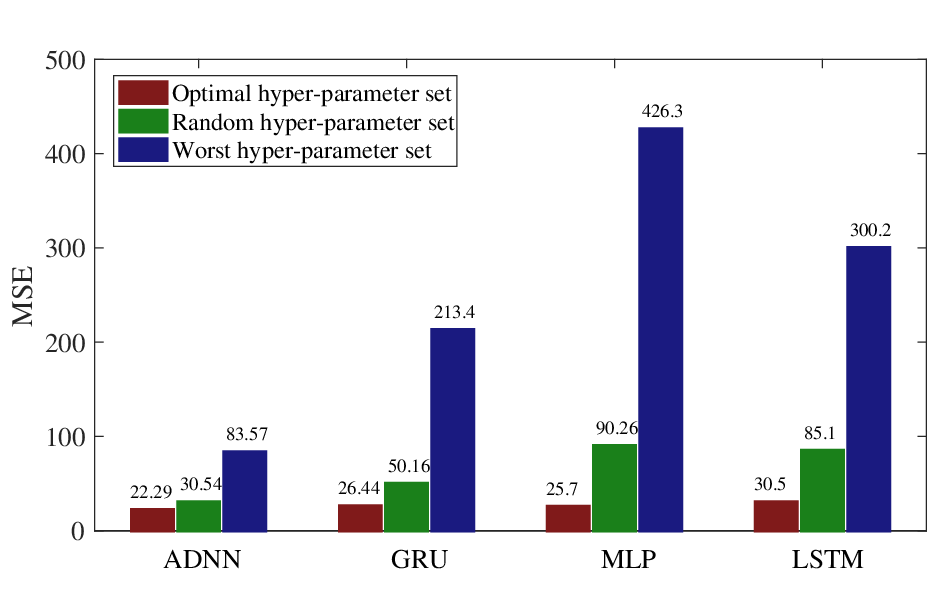}
		\end{minipage}
	}
	\centering
\vspace{-2mm}
	\caption{The mean square error (MSE) performance of prediction models with different hyper-parameter selection strategies for three randomly selected mobile cells: a) mobile cell 1595; b) mobile cell 2535; c) mobile cell 3040.}
	\label{fig1}
\vspace{-2mm}
\end{figure*}

\subsection{Preliminary analyses of hyper-parameter selection}
This paper defines the wireless NTP task as forecasting a mobile cell's future traffic load based on historical patterns observed over consecutive time intervals, with the observation window length designated as the step number $N_S$. 
We constructed prediction samples for cell $p$ using a sliding window of length $N_S$ to partition its normalized traffic series, with each windowed segment labeled by the subsequent time interval's traffic load.

We first employed four representative deep learning-based algorithms—multi-layer perceptron (MLP), LSTM, gated recurrent unit (GRU), and ADNN—to construct prediction models for each wireless NTP task and evaluated the impact of hyper-parameter selection strategies on their post-training performance.
For different algorithms, the types of hyper-parameters and their selection ranges are detailed in Table \ref{tab1} of Section V-B. Fig. \ref{fig1} presents the mean squared error (MSE) values of well-trained prediction models for three wireless NTP tasks (cells 1595, 2535, and 3040) on their testing samples, considering different algorithms and hyper-parameter selection strategies. In Fig. \ref{fig1}, the best and worst hyper-parameter selection strategies for each model are determined through grid search (GS). Then we derived the following observations.

\textbf{Observation 1}: ADNN demonstrates superior predictive potential. When adopting the optimal hyper-parameters, the ADNN-based prediction model outperforms those built with other learning algorithms in terms of accuracy. Furthermore, the hyper-parameter selection strategy plays a crucial role in determining the prediction model's capacity to achieve its optimal predictive performance.

\begin{figure}[!t]
\centerline{\includegraphics[width= 3.4in]{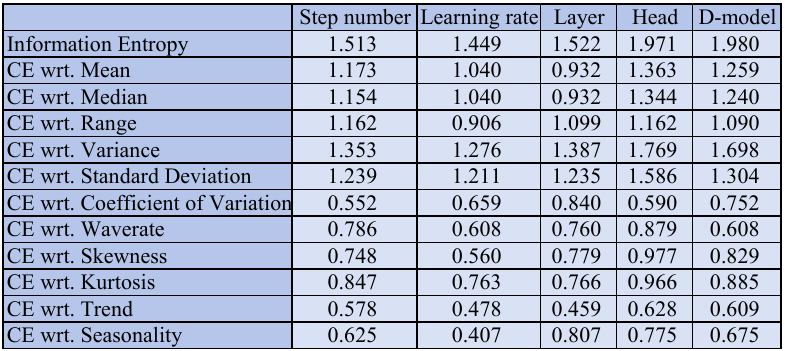}}
\caption{The information entropy and the conditional entropy (CE) of the best values of each kind of hyper-parameters.}
\label{Fig2}
\vspace{-2mm}
\end{figure}

To effectively learn, accumulate, and transfer hyper-parameter optimization experience across different NTP tasks, it is essential to define a set of features that characterize and differentiate these tasks. Moreover, the stronger the correlation between these features and the optimization target, i.e., the optimal hyper-parameters, the more feasible it becomes to establish a mapping from task features to optimal optimization results.
Therefore, we further investigated the correlation between the intrinsic characteristics or properties of NTP tasks and the optimal hyper-parameter selection strategies. Specifically, we employed ADNN as the prediction model to perform NTP tasks across different cells and then determined the optimal hyper-parameter combination for each model through GS. Then, we computed the information entropy associated with each type of hyper-parameter. Next, we selected $11$ intrinsic characteristics (e.g., mean, variance, median, etc.; see Fig. \ref{Fig2} for details) and calculated the conditional entropy of each type of hyper-parameter given each characteristic (the results are shown in Fig. \ref{Fig2}). By comparing the information entropy and conditional entropy, we derived the following observations:

\textbf{Observation 2}: 
Compared to the information entropy of optimal hyper-parameters for each class, the conditional entropy of optimal hyper-parameters given each intrinsic characteristic or property exhibits entropy reduction. This suggests a strong correlation between these intrinsic characteristics and the optimal hyper-parameters, and wireless NTP tasks with similar intrinsic characteristics or properties tend to favor similar hyper-parameter selection strategies.

\textbf{Observation 3}: Different kinds of intrinsic characteristics or properties seem to have diverse importance in the distribution of the optimal hyper-parameter selection strategies of wireless NTP models.

\section{The proposed hyper-parameter optimization framework for wireless NTP models}
This section presents an overview of the proposed meta-learning based hyper-parameter optimization framework for wireless NTP models, followed by a formal definition of its core components: 1) the base-learner, which implements fundamental NTP modeling capabilities; and 2) the meta-learner, which adaptively optimizes hyper-parameters through meta-knowledge across diverse cellular traffic patterns.

%\vspace{-2mm}
\subsection{Framework overview}
\begin{figure*}[!t]
\vspace{-0.5cm} 
\centerline{\includegraphics[width=0.75\textwidth]{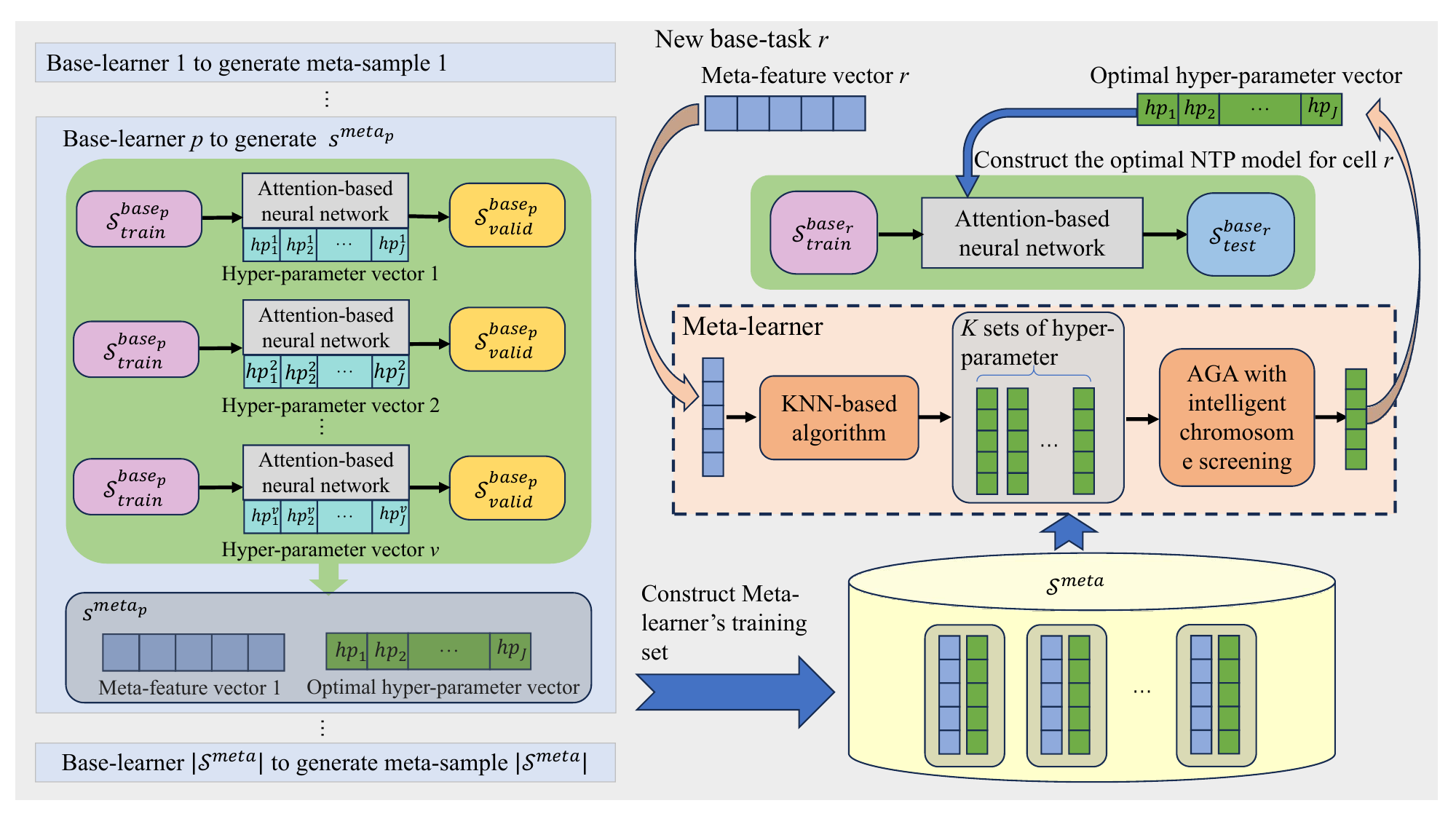}}
\caption{The proposed hyper-parameter optimization framework for cell-level wireless NTP model.}
\label{Fig3}
%\vspace{-2mm}
\end{figure*}

Fig. \ref{Fig3} illustrates a diagram of the proposed framework. In this framework, each wireless NTP task is regarded as a base-task, with ADNN-based prediction models serving as base-learners to handle these tasks. The types and ranges of hyper-parameters for base-learners are detailed in Table \ref{tab1}. 
According to meta-learning theory \cite{Meta-learning P4 right}, the hyper-parameter selection strategies collectively define the hypothesis space of a base-learner—i.e., the set of all hypothesis functions the model can represent—and significantly influence the ability to identify an optimal hypothesis during training. 
A base-learner can achieve high post-training prediction accuracy for a given base-task if: 1) Its hypothesis space contains functions that can closely approximate the target function fitting the learning samples; 2) An appropriate hypothesis function can be efficiently reached during training. 
Therefore, hyper-parameter selection is pivotal to the performance of base learners. 
Moreover, as different base-tasks correspond to varying target functions, their respective base-learners necessitate distinct hyper-parameters.

Based on Fig. \ref{Fig2} and \textbf{Observations 2 and 3}, we select $I$ intrinsic characteristics that exert the greatest influence on the distribution of optimal hyper-parameter selection strategies—i.e., those exhibiting the lowest conditional entropy across various hyper-parameter types—and designate them as the meta-features of base-tasks. 
In the proposed framework, the learning task of determining the optimal hyper-parameter selection strategy for each base-learner, based on the meta-features of its corresponding base-task, is defined as the meta-task, which is addressed by the meta-learner. 
To support the meta-learner, we construct a set of meta-samples that provides meta-knowledge. The key notations used in the proposed framework are summarized as follows.

\textbf{Notation Definitions in the Proposed Framework}: ${\cal{S}}^{meta}$ denotes the set of accumulated meta-samples. For simplicity, we also use ${\cal{S}}^{meta}$ to denote the set of base-tasks to construct the meta-samples. $s^{{meta}_p}$ in ${\cal{S}}^{meta}$ denotes the meta-sample constructed from base-task $p$ related to the $p$-th cell. ${\cal{S}}_{train}^{base_p}$ denotes the training set of base-samples for base-task $p$, while ${\cal{S}}_{valid}^{base_p}$ is base-task $p$'s validation set of base-samples to evaluate the effectiveness of different hyper-parameter selection strategies for base-learner $p$. For base-task $r \notin {\cal{S}}^{meta}$, the testing set ${\cal{S}}_{test}^{base_r}$, is constructed to evaluate its base-learner's post-training performance.

\vspace{-2mm}
\subsection{The base-learners}
Based on \textbf{Observation 1}, which highlights the predictive potential of ADNN, the base-learner for each base-task is designed as an ADNN, as illustrated in Fig. \ref{Fig4}. The ADNN comprises an encoder and a decoder, which is designed as follows.

\subsubsection{The encoder}
The encoder of a base-learner consists of $L_e$ sequential encoder blocks with an identical structure \cite{Transformer P5 left}. The input to the first encoder block is an $N_S \times D_{model}$ matrix, where the $N_S$ row vectors represent a base-task's traffic loads generated in $N_S$ continuous time intervals and $D_{model}$ denotes the dimension of the hidden layers in both encoder and decoder blocks. In the Position Encoding layer, an additional dimension of size $D_{model}$ is incorporated into the input matrix. Each encoder block consists of two sub-blocks. The first sub-block is the Multi-Head Self-Attention layer. This layer contains $H_e$ Self-Attention structures, where the key ($\cal{K}$), query ($\cal{Q}$), and value ($\cal{V}$) matrices in each structure are calculated as follows.

\begin{small}
	\begin{equation}
	\begin{array}{l}
	{\cal{K}} = {\cal{E}}_{in}^{(l_e)} \cdot {\cal{W}_{\cal{K}}},\\
	{\cal{Q}} = {\cal{E}}_{in}^{(l_e)} \cdot {\cal{W}_{\cal{Q}}},\\
	{\cal{V}} = {\cal{E}}_{in}^{(l_e)} \cdot {\cal{W}_{\cal{V}}},
	\end{array}
	\label{eq:eq109}
	\end{equation}
\end{small}
\vspace{-0.10cm}

\noindent
where ${\cal{E}}_{in}^{(l_e)}$ is the input matrix of the $l_e$-th encoder block, ${\cal{W}_{\cal{K}}}$, ${\cal{W}_{\cal{Q}}}$, and ${\cal{W}_{\cal{V}}}$ represent the learnable parameter matrices of identical dimensions \cite{Transformer P5 left}. 
Fig. \ref{Fig5} shows the structure of the Multi-Head Attention mechanism. The second sub-block is the feed-forward network with two fully-connected layers. The activation functions used in these layers are the ReLU function for the first layer and the Linear function for the second. Each sub-block is followed sequentially by a residual connection and batch normalization. The output of each encoder block is a matrix with the size of $N_S \times D_{model}$, which serves as the input to the subsequent encoder block. Specifically, the output matrix of the final encoder block, denoted as ${\cal{E}}_{out}^{(L_e)}$, is referred to as the encoded matrix.

\subsubsection{The decoder}
\begin{figure}[!t]
\vspace{-2mm}
\centerline{\includegraphics[width= 0.35\textwidth]{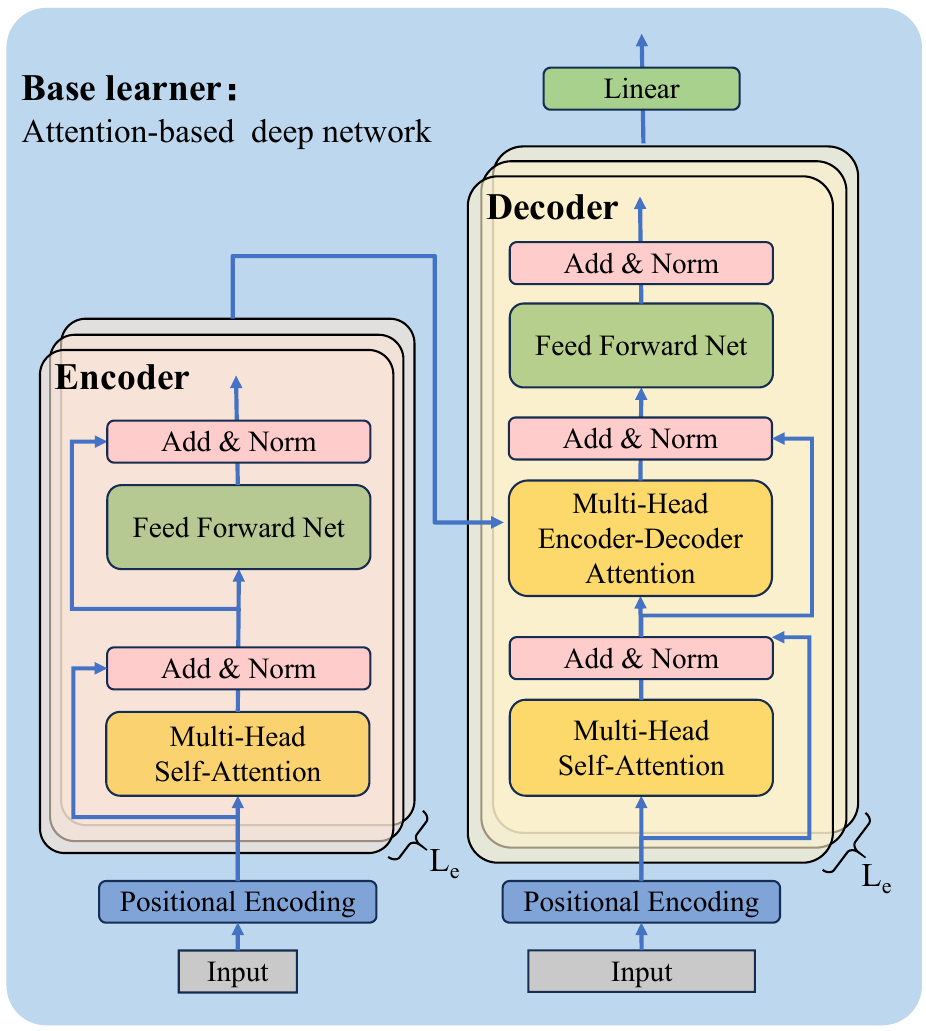}}
\caption{The structure of the base-learner.}
\label{Fig4}
\vspace{-2mm}
\end{figure}

\begin{figure}[!t]
\vspace{-2mm}
\centerline{\includegraphics[width= 1.8in]{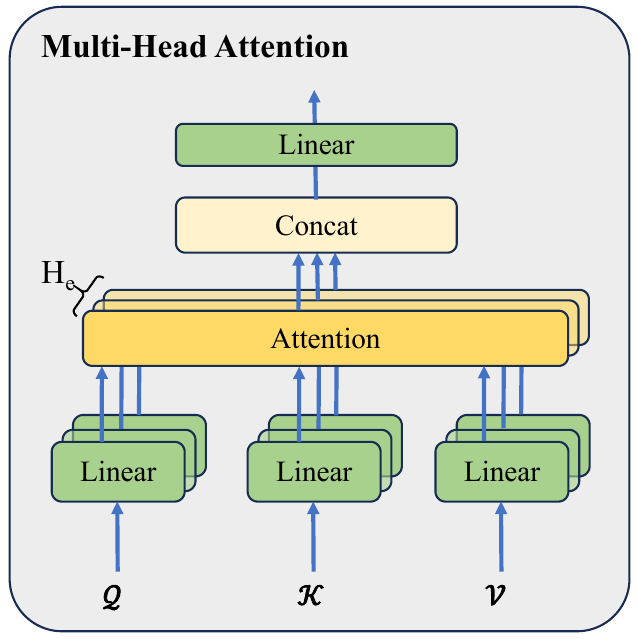}}
\caption{The schematic diagram of the Multi-Head Self-Attention mechanism.}
\label{Fig5}
\vspace{-2mm}
\end{figure}

As shown in Fig. \ref{Fig4}, the decoder consists of $L_e$ identical decoder blocks \cite{Transformer P5 left}. Each decoder block in our meta-learner comprises three sub-blocks: a Multi-Head Self-Attention layer, a Multi-Head Encoder-Decoder Attention layer, and a feed-forward network, as shown in Fig. \ref{Fig4}. The Multi-Head Self-Attention layer receives the same input matrix as the encoder and consists of $H_e$ self-attention mechanisms. Each of these mechanisms processes the input matrix using the same method as the self-attention mechanism within an encoder block. 
Let ${\cal{D}}$ denote the output matrix of the Multi-Head Self-Attention layer, which retains the same dimension as the encoder input matrix.
The Multi-Head Encoder-Decoder Attention layer takes both ${\cal{E}}_{out}^{(L_e)}$ and ${\cal{D}}$ as inputs. 
Its structure is analogous to that of the Multi-Head Self-Attention layer.
However, in this layer, the query ($\cal{Q}$) matrix is derived from ${\cal{D}}$, whereas the key ($\cal{K}$) and value ($\cal{V}$) matrices are obtained from ${\cal{E}}_{out}^{(L_e)}$ as defined by the following equations.

%\vspace{-0.5mm}
\begin{small}
	\begin{equation}
	\begin{array}{l}
	{\cal{K}} = {\cal{E}}_{out}^{(L_e)} \cdot {\cal{W}_{\cal{K}}},\\
	{\cal{Q}} = {\cal{D}} \cdot {\cal{W}_{\cal{Q}}},\\
	{\cal{V}} = {\cal{E}}_{out}^{(L_e)} \cdot {\cal{W}_{\cal{V}}}.
	\end{array}
	\label{eq:eq110}
	\end{equation}
\end{small}
\vspace{-0.10cm}

\noindent
Finally, the feed-forward network follows the same structure as the one in an encoder block. Moreover, each sub-block within the decoder block is followed by a residual connection and Batch Normalization.
The fully-connected layer then transforms the output matrix of the decoder block into a vector
where the neurons utilize the ReLU function as their activation function. The output layer of the decoder consists of a single neuron, which represents the predicted traffic load of the mobile cell for the next time interval.

For each base-learner, the hyperparameters include $L_e$, $N_S$, $D_{model}$, $H_e$, and the learning rate used in the training process, denoted as $c$.

%\vspace{-2mm}
\subsection{The proposed meta-learner}
Based on \textbf{Observation 2}, which indicates that base-tasks with similar meta-features tend to prefer similar hyper-parameters, we design a two-stage meta-learner as shown in Fig. \ref{Fig3}. In the first stage, the meta-learner utilizes KNN to identify a set of high-quality candidate hyper-parameter selection strategies for the base-learner of a newly considered base-task. 
In the second stage, the meta-learner refines the selection by applying an advanced genetic algorithm (AGA) with an intelligent chromosome screening module to determine the best hyper-parameter selection strategy.

\vspace{0.12cm}
\subsubsection{The KNN learning algorithm}
In order to leverage the KNN learning algorithm, a set of meta-samples, ${\cal{S}}^{meta}$, is constructed by randomly selecting $\left| {{S^{meta}}} \right|$ base-tasks. $\left| {{S^{meta}}} \right|$ is the cardinality of set ${\cal{S}}^{meta}$. Each meta-sample $s^{{meta}_p}$ in ${\cal{S}}^{meta}$ is constructed by labeling base-task $p$'s meta-features with the optimal hyper-parameter selection strategy of its base-learner, which is determined through a GS method.

Since different meta-features contribute unequally to hyper-parameter selection, directly using the Euclidean distance between feature vectors to represent the distance from base-task $p$ to base-task $r$ is inappropriate.
Therefore, an MLP network is employed to process meta-feature vectors of $s^{{meta}_r}$ and $s^{{meta}_p}$ through both linear and nonlinear transformations to compute the distance $D_{p-to-r}$ between $s^{{meta}_r}$ and $s^{{meta}_p}$. While the ground-truth distance $RD_{p-to-r}$ is defined as the performance obtained when $s^{{meta}_r}$ adopts the meta-label vector of $s^{{meta}_p}$. Finally, the MLP network is trained by minimizing the discrepancy between $D_{p-to-r}$ and $RD_{p-to-r}$.

For a newly considered base-task $r$, the KNN learning algorithm finds $K$ meta-samples from ${\cal{S}}^{meta}$, whose meta-feature vectors are with the shortest distances with that of base-task $r$. Since the $K$ hyper-parameter selection strategies related to the $K$ picked meta-samples are expected to provide high-quality post-training performance for base-learner $r$, they will be regarded as candidate hyper-parameter selection strategies for base-learner $r$, which will also be set as first-generation chromosomes in the following AGA.

\vspace{0.12cm}
\subsubsection{The AGA with intelligent deep learning assisted chromosome screening}
Obviously, the hyper-parameter selection space for each base-learner is quite huge. Moreover, it is almost impossible to establish a close-form mapping between the base-learner's post-training performance and the hyper-parameter selection strategy. Inspired by the fact that GA can efficiently search the solution spaces of complex optimization problems and requires no information about the forms of objective functions \cite{Genetic algorithm P6 right}, we propose an AGA to finally find the optimal hyper-parameter selection strategy for base-learner of the newly considered base-task $r \notin {\cal{S}}^{meta}$ (base-learner $r$).

Specifically, the AGA regards each kind of hyper-parameters as a fragment of one chromosome (gene) and regards each possible hyper-parameter selection strategy of a base-learner as one chromosome. 
The fitness value of a chromosome is defined as the reciprocal of base-learner $r$'s generalization error over ${\cal{S}}_{valid}^{base_r}$ when base-learner $r$ adopts the hyper-parameter selection strategy provided by this chromosome and is well-trained with ${\cal{S}}_{train}^{base_r}$. 
Besides the $K$ hyper-parameter selection strategies generated by the KNN learning algorithm, the AGA also generates $M-K$ chromosomes, in each of which the value of any gene is randomly assigned over the corresponding hyper-parameter's selection range with a uniform distribution, as its first-generation chromosomes.

\begin{figure*}[!t]
\vspace{-0.5cm} 
\subfigure[]{\begin{minipage}{0.3\linewidth}
	\flushright
    \includegraphics[width=0.75\textwidth]{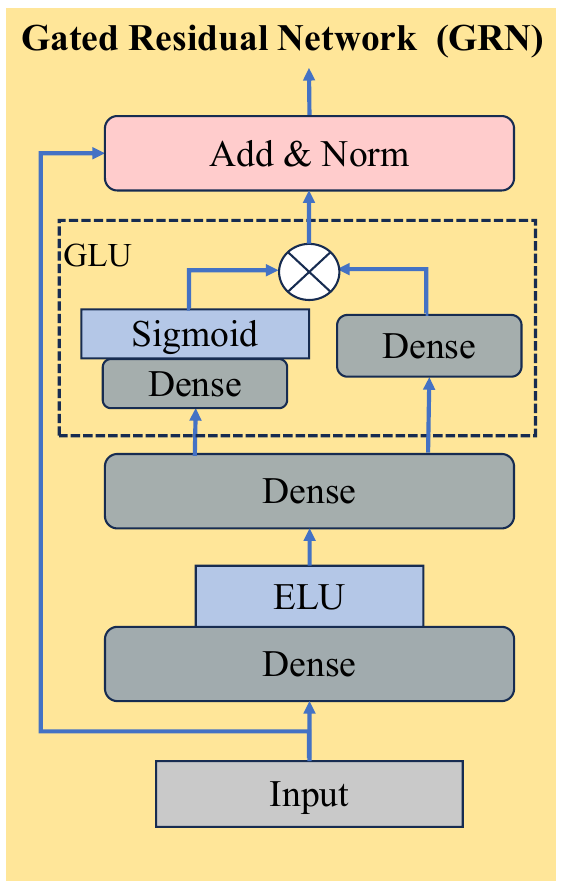}
	\flushright
\end{minipage}}
\hfill  
\subfigure[]{\begin{minipage}{0.7\textwidth}
	\centering
    \includegraphics[width=0.75\textwidth]{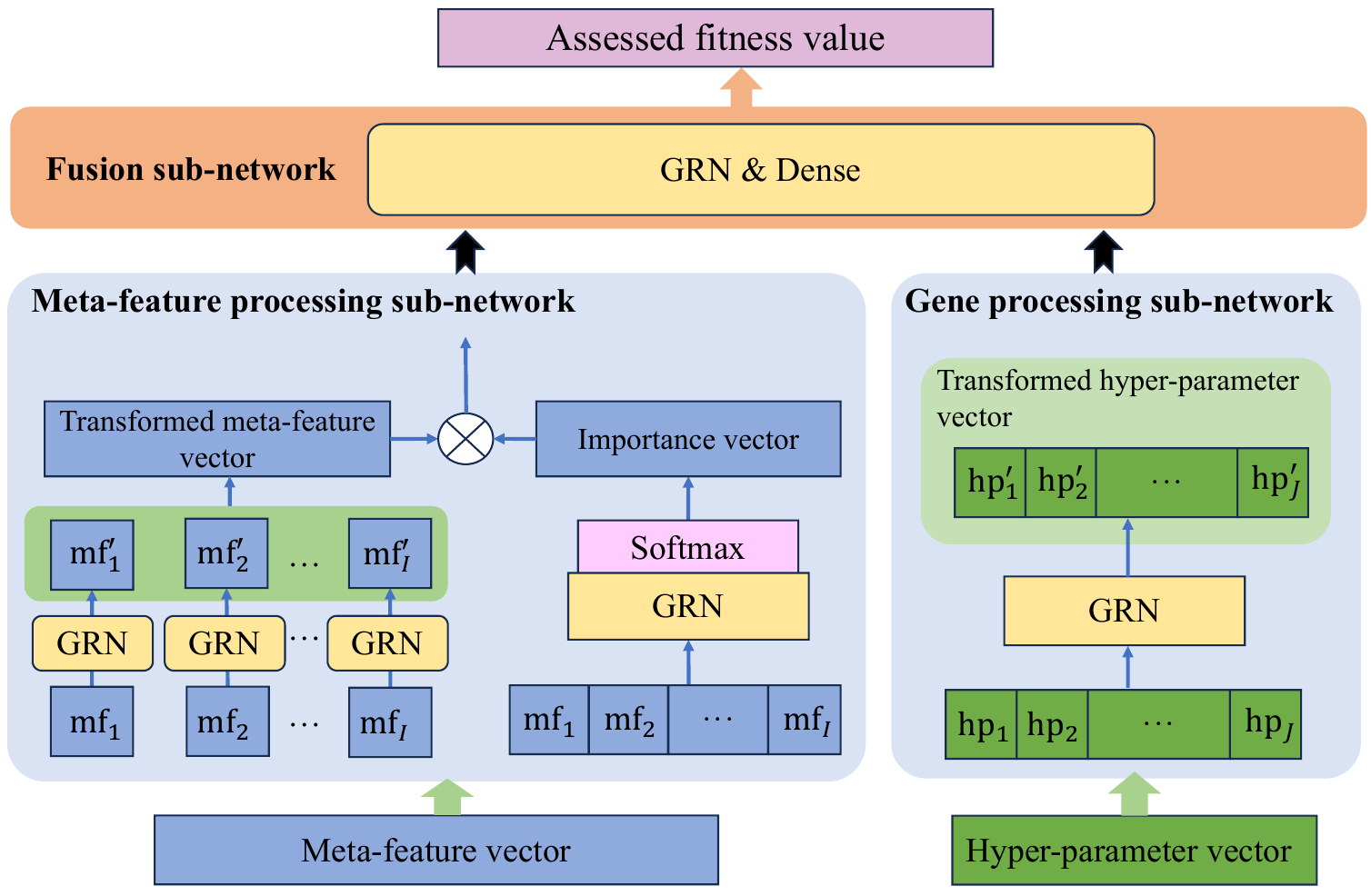}
\end{minipage}}
\caption{a) The structure of GRN; \; \; \; \; \; \; \; \; \; \; \; \; \; \; \; \; b) The framework of the chromosome screening module.}
\label{Fig6}
\vspace{-2mm}
\end{figure*}

The AGA reaches its final solution through $N$ generations of chromosomes, and there will be $M$ chromosomes surviving in each generation. The $M$ remaining chromosomes in the $n$-th ($n=1,...N-1$) generation are denoted as $\zeta _1^{\left( n \right)}$, ..., $\zeta _M^{\left( n \right)}$, whose fitness values are $f_1^{\left( n \right)}$, ..., $f_M^{\left( n \right)}$, respectively. Based on these $M$ chromosomes, $W$ ($W>>M$) son chromosomes are built. In these $W$ son chromosomes, $p_{rem} \cdot W$ ones are obtained by directly duplicating $p_{rem} \cdot W$ parent chromosomes from $\zeta _1^{\left( n \right)}$, ..., $\zeta _M^{\left( n \right)}$ with the highest fitness values. The other $(1-p_{rem}) \cdot W$ son chromosomes are hybrid ones, genes of each of which are inherited and crossed from two parent chromosomes selected randomly from $\zeta _1^{\left( n \right)}$, ..., $\zeta _M^{\left( n \right)}$. To prevent the algorithm from falling into a local optimum, any gene of every hybrid son chromosome possesses a probability of $p_{mut}$ to mutate into a random value in its selection range with a uniform distribution. 
In the proposed genetic algorithm, each parent chromosome has an equal probability of being selected to contribute to the formation of a hybrid offspring chromosome.

To mitigate the computational complexity and time consumption associated with evaluating the fitness of $W$ offspring chromosomes in each generation—requiring the assessment of base-learner $r$ under $W$ different hyper-parameter combinations—an intelligent deep learning-assisted chromosome screening module is proposed. This module accelerates the AGA in screening the surviving chromosomes in the $(n+1)$-th generation.
The AGA utilizes a sophisticated GRN-based chromosome screening module as shown in Fig. \ref{Fig6} to evaluate fitness values of the $W$ son chromosomes and selects $\tau \cdot M$ son chromosomes with the highest evaluated fitness values, where $\tau$ is a constant larger than 1. After that, fitness values of only those $\tau \cdot M$ son chromosomes are calculated, and the $M$ ones with the largest actual fitness values survive as the next-generation chromosomes. Please note that since $\tau \cdot M \ll W$, the novel intelligent deep learning-assisted chromosome screening scheme will improve the computational efficiency of the AGA tremendously.

As shown in Fig. \ref{Fig6} (b), the proposed chromosome screening module takes base-task $r$'s meta-features and hyper-parameter selection strategy related to a son chromosome as its inputs while outputs the evaluated fitness value for this son chromosome. With the motivation of giving the deep neural network flexibility to apply linear processing and non-linear processing of its inputs only where needed, the GRN structure is presented in Fig. \ref{Fig6} (a) as a building block of the deep neural network. A GRN block $\omega$ takes in an input vector ${\bf{i}}$ and yields

%\vspace{-0.20cm}
\begin{small}
	\begin{equation}
	GRN_\omega \left( {\bf{i}} \right) = LayerNorm\left( {{\bf{i}} + GLU_\omega \left( {{\eta _1}} \right)} \right),
	\label{eq:eq112}
	\end{equation}
\end{small}
\vspace{-0.80cm}

\begin{small}
	\begin{equation}
	{\eta _1} = {{\bf{O}}_{1,\omega }}{\eta _2} + {{\bf{b}}_{1,\omega }},
	\label{eq:eq113}
	\end{equation}
\end{small}
\vspace{-0.80cm}

\begin{small}
	\begin{equation}
	{\eta _2} = ELU\left( {{{\bf{O}}_{2,\omega }}{\bf{i}} + {{\bf{b}}_{2,\omega }}} \right),
	\label{eq:eq114}
	\end{equation}
\end{small}
\vspace{-0.20cm}

\noindent
where ELU is the exponential linear unit activation function \cite{GRN P7 left}; LayerNorm is a standard normalization layer; $\eta _1$ and $\eta _2$ are intermediate layer outputs. Taking $\eta _1$ as the input, the gated linear units (GLU) module generates

%\vspace{-0.01cm}
\begin{small}
	\begin{equation}
	GL{U_\omega }\left( {{\eta _1}} \right) = \sigma \left( {{{\bf{O}}_{3,\omega }}{\eta _1} + {{\bf{b}}_{3,\omega }}} \right) \odot \left( {{{\bf{O}}_{4,\omega }}{\eta _1} + {{\bf{b}}_{4,\omega }}} \right),
	\label{eq:eq115}
	\end{equation}
\end{small}
\vspace{-0.20cm}

\noindent
where $\sigma \left(  \cdot  \right)$ is the sigmoid activation function, $ \odot $ is the element-wise Hadamard product. In equations (5-7), ${{\bf{O}}_{\left(  \cdot  \right)}}$ and ${{\bf{b}}_{\left(  \cdot  \right)}}$ are neuron connection weight matrix and neuron bias vector, respectively. The GLU module allows a GRN block to control the extent to which non-linear processing of the input vector ${\bf{i}}$ contributes to the output vector, e.g., the GLU outputs could be close to $0$ to suppress the nonlinear contribution.

The GRN-based chromosome screening module demonstrated in Fig. \ref{Fig6} (b) contains three sub-networks: the meta-feature processing (MFP) sub-network, the gene processing (GP) sub-network, and the fusion sub-network. 
The MFP sub-network takes base-task $r$'s meta-features as its input vector and applies linear and non-linear processing for the inputs via GRN blocks, generating the transformed meta-features. 
Due to the fact that different kinds of meta-features have diverse ranging scales and contribute unequally to base-learner $r$'s performance, an automatic importance evaluation mechanism is designed for the meta-features in this sub-network. Specifically, taking base-task $r$'s meta-features as the inputs, an importance vector is generated through a GRN block and a Softmax layer. The MFP sub-network then takes the Hadamard product between the transformed meta-features and the importance vector as its outputs. 
The GP sub-network takes the genes determined by a son chromosome as its inputs and yields the transformed gene values via multiple GRN blocks. Finally, the fusion sub-network conducts linear and non-linear processing for the outputs of the former two sub-networks via a GRN block and yields the evaluated fitness values of a son chromosome for base-learner $r$ through a full-connection neural network.

The MFP sub-network and the GP sub-network each utilize a two-layer fully connected network (512 hidden units, ReLU activation), incorporating residual connections and layer normalization. The fusion sub-network integrates meta-feature importance weighting and hyper-parameter fusion, followed by a three-layer fully connected network for final output. Training employs a learning rate of 0.001, a batch size of 252, and 400 epochs, ensuring stable optimization and effective feature integration.

The chromosome screening module is trained with learning samples generated from base-tasks in ${\cal{S}}^{meta}$. For any base-task $p \in {\cal{S}}^{meta}$, since the fitness values of all the possible hyper-parameter selection strategies have been tested for base-learner $p$, numerous learning samples can be built related to this base-task for the chromosome screening module by labeling each hyper-parameter selection strategy as well as base-task $p$'s meta-features with the corresponding fitness value.

Finally, the pseudo-code is presented in \textbf{Algorithm 1} to illustrate the meta-learner's operating process in detail.

%\vspace{0.2cm} 
\begin{algorithm}
\label{algorithm1}
	\renewcommand{\algorithmicrequire}{\textbf{Input:}}
	\renewcommand{\algorithmicensure}{\textbf{Output:}}
	\caption{}
	\begin{algorithmic}[1]
	\REQUIRE ${\cal{S}}^{meta}, p_{rem}, p_{mut}, \text{base-task}~r $ 
		\STATE From the $r$-th traffic prediction task, i.e. base-task $r$, obtain the corresponding meta-feature vector;
		\STATE Obtain $K$ neighboring meta-samples from ${\cal{S}}^{meta}$ whose meta-feature vectors have the smallest distances with the meta-feature vector of base-task $r$ with the KNN learning algorithm;
		\STATE Generate $K$ first-generation chromosomes ($K$ hyper-parameter selection strategies for base-learner $r$) with the labels of the $K$ selected meta-samples;
		\STATE Generate $M-K$ first-generation chromosomes randomly;
		\STATE Construct the set of $M$ first-generation chromosomes based on the outputs of steps 3 and 4;
		\STATE Calculate the fitness value of each first-generation chromosome; 
		\FOR{$n = 1:N$}
		\STATE Select $p_{rem} \cdot M$ chromosomes in the $n$-th generation ones with the largest fitness values as the son chromosomes;
			\FOR{$w = 1:(W-p_{rem} \cdot M)$}
			\STATE Select two chromosomes in the $n$-th generation ones randomly with the uniform selecting probability;
			\STATE Generate a son chromosome by inheriting and crossing the genes of the two selected chromosomes;
			\STATE Mutate each gene of the son chromosome with a probability of $p_{mut}$ into a random value in the gene's selection range;
			\ENDFOR
		\STATE Construct the set of $W$ son chromosomes for the $n$-th generation ones based on the outputs of steps 10-12;  
		\STATE Evaluate the fitness values of the $W$ son chromosomes with the proposed chromosome screening module;
		\STATE Select $\tau \cdot M$ son chromosomes with the largest evaluated fitness values;
		\STATE Calculate the practical fitness values of $\tau \cdot M$ son chromosomes and obtain $M$ son chromosomes with the largest fitness values as the survived ones ($n+1$)-th generation chromosome;
		\ENDFOR
	\ENSURE The chromosome having the largest fitness value within $N$ generations of chromosomes
	\end{algorithmic}  
\end{algorithm}

\vspace{-0.2cm}
\section{Numerical Results of the Proposed Framework}
This section first introduces the experimental settings and performance metrics. Subsequently, the effectiveness of the KNN learning algorithm in the proposed meta-learner is numerically validated, and the impact of key parameters on the performance of the AGA is analyzed. Next, the prediction accuracy and computational complexity of the proposed framework are compared with several benchmark methods. Finally, the robustness of \textbf{Algorithm 1} is assessed.

%\vspace{-0.2cm}
\subsection{Experimental settings}
In the adopted dataset, traffic load records for certain mobile cells are missing during some time intervals due to collection failures or storage errors. To address this, a widely used method \cite{Average traffic load P8 left} is applied, where each missing value is replaced with the average traffic load of the target cell's eight surrounding cells during the same time interval.

\begin{table*}
\vspace{-0.5cm} 
\setlength{\belowcaptionskip}{-0.7cm} 
\begin{center}
\caption{The hyper-parameters' selection range.}
\label{tab1}
\begin{tabular}{| c | c | c | c | c | c |}
\hline
 & Step number ($N_S$) & Learning rate ($c$) & Layer ($L_e$) & Head ($H_e$) & ${D_{model}}$ \\
\hline
ADNN & (6, 12, 18) & (0.01, 0.001, 0.0001) & (1, 2, 3) & (2, 4, 6, 8) & (8, 16, 32, 64, 128, 256, 512)\\
\hline
 & Step number & Learning rate & Layer & \multicolumn{2}{c|}{Neure} \\
\hline
GRU & (6, 12, 18) & (0.01, 0.001, 0.0001) & (2, 3, 4) & \multicolumn{2}{c|}{(256, 512, 768)} \\
\hline
LSTM & (6, 12, 18) & (0.01, 0.001, 0.0001) & (2, 3, 4) & \multicolumn{2}{c|}{(32, 64, 128, 256)} \\
\hline 
MLP & (6, 12, 18) & (0.01, 0.001, 0.0001) & (2, 3, 4) & \multicolumn{2}{c|}{(128, 256, 512)} \\
\hline 
\end{tabular}
\end{center}
\vspace{-8mm}
\end{table*}

The performance of the base-learner, ADNN, is evaluated against several representative wireless NTP models, including Support Vector Regression (SVR), Gaussian Process (GP), and conventional deep learning-based methods such as MLP, LSTM, and GRU, as shown in Table \ref{tab2}.
To evaluate how the proposed framework enhances the post-training performance of the base-learner by providing the optimal hyper-parameters, conventional deep learning-based methods are tested with randomly selected hyper-parameters within the selection ranges listed in Table \ref{tab1}. 
For ADNN-based base-learners, their performances under different hyper-parameter optimization methods are further evaluated, including stochastic search (SS), GA, BO, particle swarm optimization (PSO), GS, and the proposed framework in Algorithm 1.
Moreover, we introduce two partial implementations of the proposed framework: (1) an AGA with deep learning-assisted chromosome screening module but randomly initialized the first-generation chromosomes; and (2) GA+KNN, where the initial population is selected using KNN but without the proposed chromosome screening module.

A subset of mobile cells is randomly selected from the dataset to form the meta-sample set ${\cal{S}}^{meta}$ for the proposed framework, with the remaining cells designated as testing base-tasks through an additional random sampling process.
For each mobile cell $p \in {\cal{S}}^{meta}$, the training and validation sets for its base-learner are constructed by samples collected from November 1, 2013, to November 30, 2013. 
Similarly, for each mobile cell $r \notin {\cal{S}}^{meta}$, the training and validation sets are also constructed using samples from November 1, 2013, to November 30, 2013, whereas the testing set consists of samples collected from December 1, 2013, to December 3, 2013.
To ensure a fair comparison, the training samples used to train models in conventional deep learning-based prediction methods for each testing task are also selected from the period of November 1, 2013, to November 30, 2013.

The base-learners in the proposed framework, along with the deep learning-based prediction models, are optimized using AdamW \cite{AdamW}, a widely adopted stochastic gradient-based optimization technique in the machine learning domain. MSE is used as the loss function during model training.
To evaluate prediction accuracy, two metrics are employed: the MSE and the coefficient of determination (R2). The R2 metric quantifies the proportion of variance in the ground truth values that is explained by the prediction model, while MSE represents the average squared difference between predicted and true values.

%\vspace{-0.2cm}
\subsection{Effectiveness of the KNN learning algorithm}
This subsection experimentally verifies the effectiveness of the KNN learning algorithm used by the meta-learner for generating high-quality first-generation chromosomes.

\begin{figure}[!t]
\setlength{\abovecaptionskip}{-0.05cm} 
\centerline{\includegraphics[width= 3.1in]{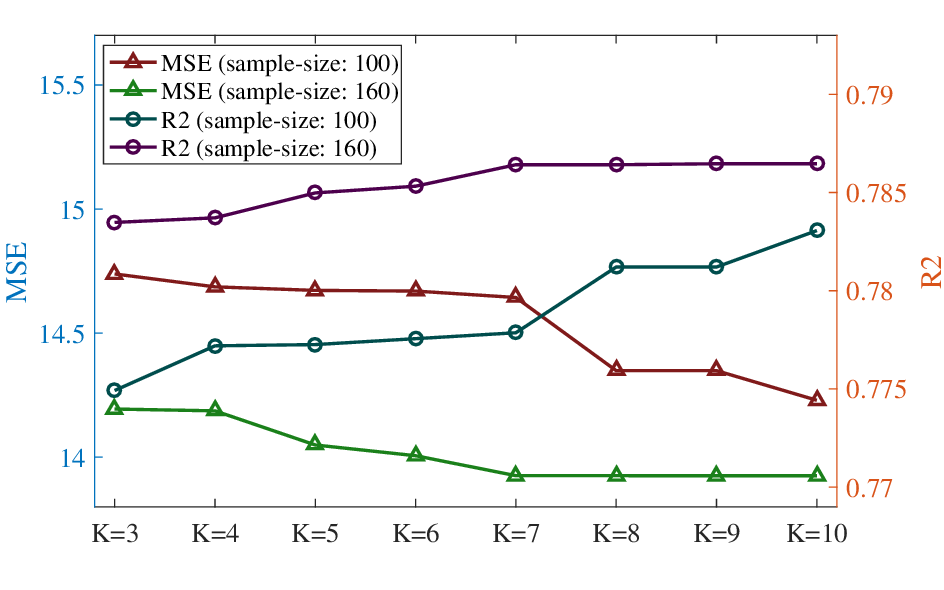}}
\caption{The average post-training MSE and R2 performance achieved by base-learners when each base-learner adopts the conditionally optimal hyper-parameter selection strategy provided by the KNN learning algorithm versus the neighbor number $K$ under different scales of ${\cal{S}}^{meta}$.}
\label{fig:MSE-Kvalue}
\vspace{-2mm}
\end{figure}

For the testing base-tasks, Fig. \ref{fig:MSE-Kvalue} illustrates the average post-training MSE and R2 performance of their base-learners over validation sets, where each base-learner adopts the best hyper-parameters from the candidate strategies provided by its neighboring meta-samples in ${\cal{S}}^{meta}$, as a function of the neighbor number K in the KNN learning algorithm under different sizes of ${\cal{S}}^{meta}$.
A lower MSE or higher R2 indicates that the KNN learning algorithm in the meta-learner effectively provides superior candidate hyper-parameter selection strategies for the base-learners in testing base-tasks. 
As shown in Fig. \ref{fig:MSE-Kvalue}, for a fixed ${\cal{S}}^{meta}$ size, the performance of the KNN learning algorithm improves noticeably as $K$ increases. This improvement is attributed to the fact that incorporating more neighboring meta-samples enhances the likelihood of selecting a high-quality hyper-parameter selection strategy from their labels, thereby improving post-training prediction accuracy. 
An interesting observation in Fig. \ref{fig:MSE-Kvalue} is that as $K$ increases from 3 to 10, the performance of the KNN learning algorithm improves rapidly at first and then stabilizes. Specifically, when $K$ exceeds 8, further increasing $K$ results in only marginal performance gains.

\begin{figure}[!t]
\setlength{\abovecaptionskip}{-0.05cm} 
\centerline{\includegraphics[width= 3.1in]{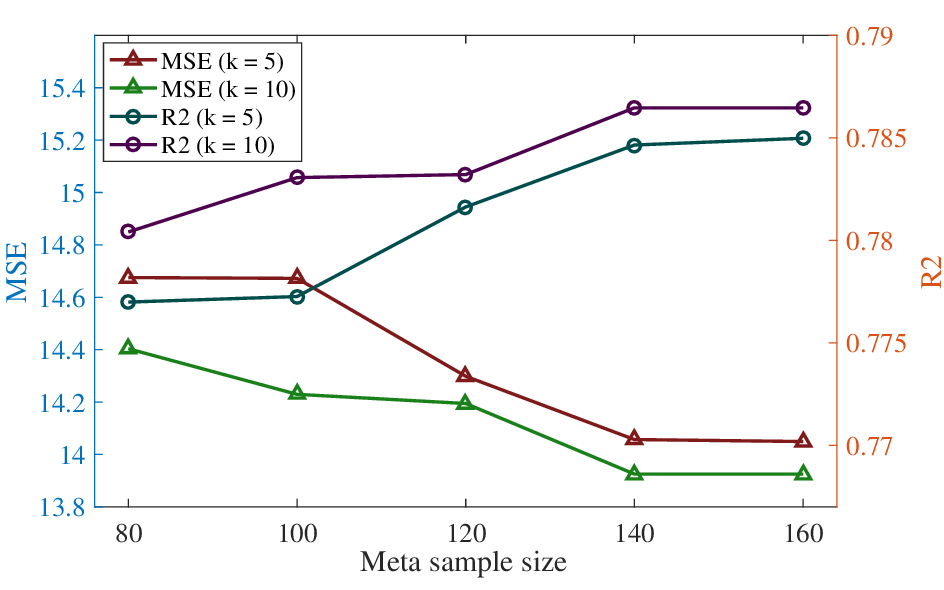}}
\caption{The average post-training MSE and R2 performance achieved by base-learners when each base-learner adopts the conditionally optimal hyper-parameter selection strategy provided by the KNN learning algorithm versus the scale of ${\cal{S}}^{meta}$ with different values of $K$.}
\label{fig:MSE-Samplesize}
\vspace{-2mm}
\end{figure}

Fig. \ref{fig:MSE-Samplesize} illustrates the performance of the KNN learning algorithm as a function of the scale of ${\cal{S}}^{meta}$ under different values of $K$. It can be observed that, for a fixed $K$, the performance of the KNN learning algorithm improves as the size of ${\cal{S}}^{meta}$ increases.
This can be explained by the fact that as more base-tasks are solved and meta-knowledge accumulates, the KNN learning algorithm becomes more likely to identify meta-samples with meta-features similar to those of a given testing base-task. Consequently, these meta-samples facilitate more effective hyper-parameter selection for the testing base-learner. It aligns with \textbf{Observation 3}, which suggests that base-tasks with similar meta-features tend to prefer similar hyper-parameters. 
Moreover, as shown in Fig. \ref{fig:MSE-Samplesize}, 
the effectiveness of the KNN learning algorithm gradually flattens when the size of ${\cal{S}}^{meta}$ exceeds 140.
This suggests that acquiring excessive meta-knowledge may not be necessary to ensure the algorithm's performance.

Throughout the experiments, the parameter $K$ and the scale of ${\cal{S}}^{meta}$ are set to $8$ and $160$, respectively, to balance the framework performance and the calculation complexity.

%\vspace{-0.2cm}
\subsection{Influence of key parameters in the AGA}
The proportion of parent chromosomes retained in the $W$ candidate offspring chromosomes $p_{rem}$, and the gene mutation probability $p_{mut}$, are two key parameters of the proposed AGA in the meta-learner. This subsection investigates the impact of these parameters on the performance of the proposed algorithm when part of the first-generation chromosomes is generated with the assistance from KNN.

\begin{figure}[!t]
\setlength{\abovecaptionskip}{0.04cm} 
\centerline{\includegraphics[width= 3.1in]{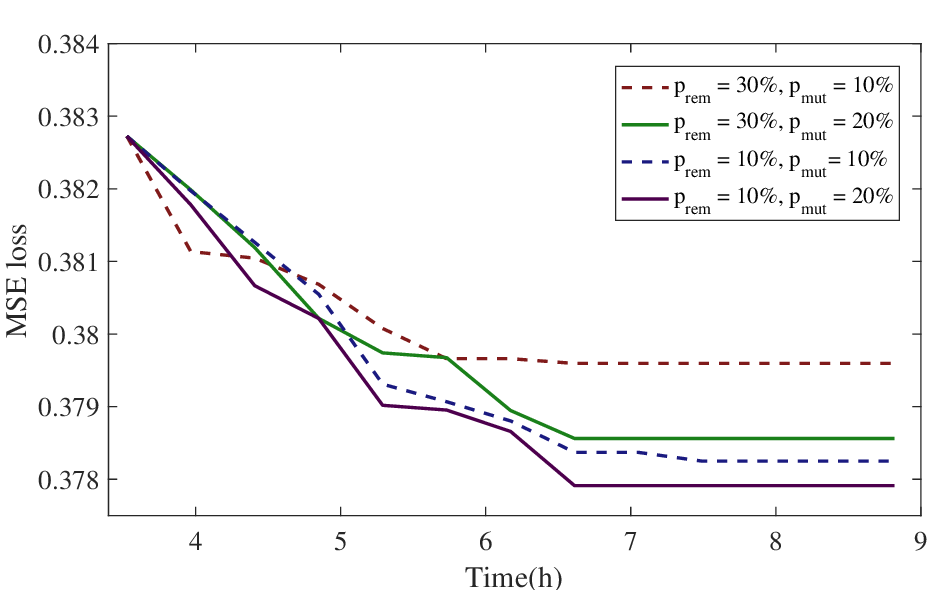}}
\caption{The base-learner's post-training MSE performance with the currently optimal hyper-parameter selection strategy output by \textbf{Algorithm 1} for mobile cell 1635 versus \textbf{Algorithm 1}'s processing time under different value combinations of $p_{rem}$ and $p_{mut}$.}
\label{fig:GA hyper-parameter 1635}
\vspace{-2mm}
\end{figure}

\begin{figure}[!t]
\setlength{\abovecaptionskip}{0.04cm} 
\centerline{\includegraphics[width= 3.1in]{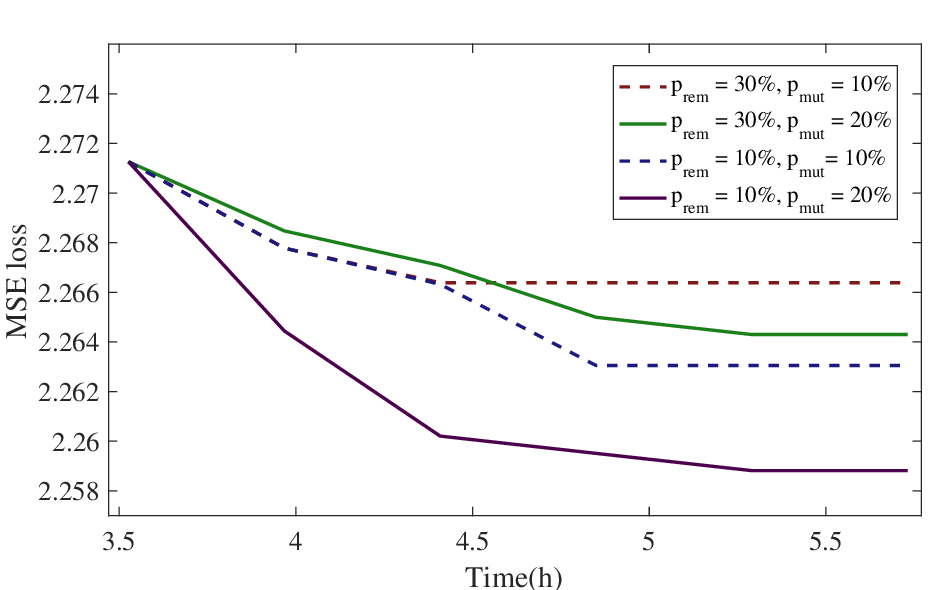}}
\caption{The base-learner's post-training MSE performance with the currently optimal hyper-parameter selection strategy output by \textbf{Algorithm 1} for mobile cell 4004 versus \textbf{Algorithm 1}'s processing time under different value combinations of $p_{rem}$ and $p_{mut}$.}
\label{fig:GA hyper-parameter 4004}
\vspace{-2mm}
\end{figure}

For two randomly selected testing base-tasks (mobile cells 1635 and 4004), Figs. \ref{fig:GA hyper-parameter 1635} and \ref{fig:GA hyper-parameter 4004} demonstrate the post-training performance of base-learners using optimal hyper-parameters obtained by the AGA, along with the corresponding time consumption of hyper-parameter optimization under specific values of $p_{rem}$ and $p_{mut}$. 
Specifically, each curve in Fig. \ref{fig:GA hyper-parameter 1635} and Fig. \ref{fig:GA hyper-parameter 4004} represents results averaged over $10$ independent runs of \textbf{Algorithm 1}. These figures indicate that the algorithm exhibits a faster convergence speed but lower post-convergence performance when $p_{rem}$ is large or $p_{mut}$ is small. 
This phenomenon can be attributed to the fact that, under such settings, a larger proportion of high-quality parent chromosomes and beneficial genes are preserved in the next generation.
As a result, the algorithm can quickly converge to a solution for the hyper-parameter optimization problem in each base-task. Moreover, since more parent chromosomes are retained in the next-generation chromosome set, the algorithm can bypass recalculating their fitness values in each iteration. This avoids the computationally expensive process of retraining the base-learner with parent chromosomes, as these fitness values have already been evaluated.
However, when $p_{rem}$ is large or $p_{mut}$ is small, the algorithm tends to rely heavily on existing chromosomes and genes, increasing the risk of getting trapped in a local optimum and failing to produce an optimal solution.
In contrast, when $p_{rem}$ is small and $p_{mut}$ is large, Figs. \ref{fig:GA hyper-parameter 1635} and \ref{fig:GA hyper-parameter 4004} show that the AGA achieves superior final performance, albeit with a slower convergence rate. 
This is because greater exploration in each iteration improves the chances of escaping local optima and finding the global optimum, at the expense of longer search time.

Regardless of the specific values of $p_{rem}$ and $p_{mut}$, Figs. \ref{fig:GA hyper-parameter 1635} and \ref{fig:GA hyper-parameter 4004} demonstrate that the proposed AGA effectively refines the hyper-parameter selection strategies initially generated by the KNN learning algorithm, while maintaining a reasonable convergence time. To further enhance post-convergence performance, a relatively small $p_{rem}$ ($10\%$) and a relatively large $p_{mut}$ ($20\%$) are adopted in the final configuration. 

%\vspace{-0.2cm}
\subsection{Prediction accuracy of the proposed framework and benchmark methods}
The post-training accuracy performance of the proposed framework is compared with several benchmark methods. Table \ref{tab2} summarizes the MSE and R2 values for the evaluated prediction methods, including results from five randomly selected testing base-tasks as well as the average performance across all considered testing base-tasks.

As shown in Table \ref{tab2}, shallow learning-based methods, namely SVR and GP, exhibit relatively low prediction accuracy despite their ability to capture nonlinearities in time series. This is primarily due to the complex temporal autocorrelations present in cell-level wireless network traffic patterns, which exceed the learning capacity of these methods. 
The limited number of tunable parameters in SVR and GP hinders their ability to capture the intricate characteristics of cell-level wireless network traffic patterns, resulting in underfitting and suboptimal predictive accuracy.
In contrast, conventional deep learning-based methods—including MLP, LSTM, GRU, and ADNN—employ more complex architectures and a large number of trainable parameters. This allows them to capture deep dependencies and dynamic variations in traffic loads across time intervals. Consequently, these methods achieve lower average MSE and higher average R2 across the testing base-tasks, outperforming SVR and GP.

However, Table \ref{tab2} also reveals that conventional deep learning models exhibit unstable post-training accuracy across different base-tasks. For example, the MLP network produces a large prediction error for base-task 6035, while ADNN underperforms for base-task 9106. This aligns with \textbf{Observation 1}, which highlights that hyper-parameter selection strategies significantly impact deep learning model performance. Poorly chosen hyper-parameters can severely degrade prediction accuracy, explaining the observed instability across base-tasks.

\begin{table*}
\vspace{-0.5cm} 
\setlength{\belowcaptionskip}{-0.7cm} 
\begin{center}
\caption{The performance of the proposed framework and the benchmark methods}
\label{tab2}
\begin{tabular}{| c | c | c | c | c | c | c | c | c | c | c | c | c |}
\hline
\multicolumn{13}{|c|}{Benchmark methods without hyper-parameter optimization}\\
\hline
 & \multicolumn{2}{c|}{cell 5824} & \multicolumn{2}{c|}{cell 6035} & \multicolumn{2}{c|}{cell 6036} & \multicolumn{2}{c|}{cell 8417} & \multicolumn{2}{c|}{cell 9106} & \multicolumn{2}{c|}{Mean} \\
\hline
 & {\fontsize{8}{11}\selectfont MSE} & {\fontsize{8}{11}\selectfont R2} & {\fontsize{8}{11}\selectfont MSE} & {\fontsize{8}{11}\selectfont R2} & {\fontsize{8}{11}\selectfont MSE} & {\fontsize{8}{11}\selectfont R2} & {\fontsize{8}{11}\selectfont MSE} & {\fontsize{8}{11}\selectfont R2} & {\fontsize{8}{11}\selectfont MSE} & {\fontsize{8}{11}\selectfont R2} & {\fontsize{8}{11}\selectfont MSE} & {\fontsize{8}{11}\selectfont R2}\\
\hline
{\fontsize{8}{11}\selectfont SVR} 
& {\fontsize{7}{11}\selectfont 19.37} & {\fontsize{7}{11}\selectfont 20.98\%} 
& {\fontsize{7}{11}\selectfont 5.12} & {\fontsize{7}{11}\selectfont 41.58\%} 
& {\fontsize{7}{11}\selectfont 61.14} & {\fontsize{7}{11}\selectfont 22.02\%} 
& {\fontsize{7}{11}\selectfont 122} & {\fontsize{7}{11}\selectfont 46.22\%} 
& {\fontsize{7}{11}\selectfont 58.62} & {\fontsize{7}{11}\selectfont 20.64\%} 
& {\fontsize{7}{11}\selectfont 36.21} & {\fontsize{7}{11}\selectfont 44.96\%} \\
\hline
{\fontsize{8}{11}\selectfont GP} 
& {\fontsize{7}{11}\selectfont 21.2} & {\fontsize{7}{11}\selectfont 13.52\%} 
& {\fontsize{7}{11}\selectfont 6.45} & {\fontsize{7}{11}\selectfont 26.34\%} 
& {\fontsize{7}{11}\selectfont 50.02} & {\fontsize{7}{11}\selectfont 36.21\%} 
& {\fontsize{7}{11}\selectfont 148.1} & {\fontsize{7}{11}\selectfont 34.71\%} 
& {\fontsize{7}{11}\selectfont 31.48} & {\fontsize{7}{11}\selectfont 57.38\%} 
& {\fontsize{7}{11}\selectfont 41.27} & {\fontsize{7}{11}\selectfont 40.37\%} \\
\hline
{\fontsize{8}{11}\selectfont GRU} 
& {\fontsize{7}{11}\selectfont 12.13} & {\fontsize{7}{11}\selectfont 50.53\%} 
& {\fontsize{7}{11}\selectfont 3.52} & {\fontsize{7}{11}\selectfont 59.86\%} 
& {\fontsize{7}{11}\selectfont 66.05} & {\fontsize{7}{11}\selectfont 15.76\%} 
& {\fontsize{7}{11}\selectfont 81.31} & {\fontsize{7}{11}\selectfont 64.16\%} 
& {\fontsize{7}{11}\selectfont 15.65} & {\fontsize{7}{11}\selectfont 78.81\%} 
& {\fontsize{7}{11}\selectfont 24.71} & {\fontsize{7}{11}\selectfont 62.6\%} \\
\hline
{\fontsize{8}{11}\selectfont LSTM} 
& {\fontsize{7}{11}\selectfont 22.47} & {\fontsize{7}{11}\selectfont 8.35\%} 
& {\fontsize{7}{11}\selectfont 5.33} & {\fontsize{7}{11}\selectfont 39.13\%} 
& {\fontsize{7}{11}\selectfont 49.24} & {\fontsize{7}{11}\selectfont 37.19\%} 
& {\fontsize{7}{11}\selectfont 96.58} & {\fontsize{7}{11}\selectfont 57.43\%} 
& {\fontsize{7}{11}\selectfont 22.66} & {\fontsize{7}{11}\selectfont 69.32\%} 
& {\fontsize{7}{11}\selectfont 28.68} & {\fontsize{7}{11}\selectfont 52.39\%} \\
\hline
{\fontsize{8}{11}\selectfont MLP} 
& {\fontsize{7}{11}\selectfont 19.44} & {\fontsize{7}{11}\selectfont 20.7\%} 
& {\fontsize{7}{11}\selectfont 5.8} & {\fontsize{7}{11}\selectfont 33.82\%} 
& {\fontsize{7}{11}\selectfont 67.83} & {\fontsize{7}{11}\selectfont 13.48\%} 
& {\fontsize{7}{11}\selectfont 78.13} & {\fontsize{7}{11}\selectfont 65.56\%} 
& {\fontsize{7}{11}\selectfont 19.04} & {\fontsize{7}{11}\selectfont 74.22\%} 
& {\fontsize{7}{11}\selectfont 26.2} & {\fontsize{7}{11}\selectfont 55.71\%} \\
\hline
{\fontsize{8}{11}\selectfont ADNN} 
& {\fontsize{7}{11}\selectfont 10.68} & {\fontsize{7}{11}\selectfont 56.45\%} 
& {\fontsize{7}{11}\selectfont 5.97} & {\fontsize{7}{11}\selectfont 31.87\%} 
& {\fontsize{7}{11}\selectfont 55.97} & {\fontsize{7}{11}\selectfont 28.61\%} 
& {\fontsize{7}{11}\selectfont 90.89} & {\fontsize{7}{11}\selectfont 59.94\%} 
& {\fontsize{7}{11}\selectfont 29.32} & {\fontsize{7}{11}\selectfont 60.31\%} 
& {\fontsize{7}{11}\selectfont 25.9} & {\fontsize{7}{11}\selectfont 59.22\%} \\
\hline
\multicolumn{13}{|c|}{Traditional Hyper-parameter Optimization for ADNN}\\
\hline
{\fontsize{8}{11}\selectfont SS} 
& {\fontsize{7}{11}\selectfont 9.51} & {\fontsize{7}{11}\selectfont 61.21\%} 
& {\fontsize{7}{11}\selectfont 3.72} & {\fontsize{7}{11}\selectfont 57.53\%} 
& {\fontsize{7}{11}\selectfont 45.6} & {\fontsize{7}{11}\selectfont 41.83\%} 
& {\fontsize{7}{11}\selectfont 81.67} & {\fontsize{7}{11}\selectfont 64\%} 
& {\fontsize{7}{11}\selectfont 16.64} & {\fontsize{7}{11}\selectfont 77.47\%} 
& {\fontsize{7}{11}\selectfont 21.62} & {\fontsize{7}{11}\selectfont 70.25\%} \\
\hline
{\fontsize{8}{11}\selectfont GA} 
& {\fontsize{7}{11}\selectfont 9.74} & {\fontsize{7}{11}\selectfont 60.26\%} 
& {\fontsize{7}{11}\selectfont 4.32} & {\fontsize{7}{11}\selectfont 50.74\%}
& {\fontsize{7}{11}\selectfont 34.8} & {\fontsize{7}{11}\selectfont 55.62\%} 
& {\fontsize{7}{11}\selectfont 84.7} & {\fontsize{7}{11}\selectfont 62.66\%} 
& {\fontsize{7}{11}\selectfont 13.8} & {\fontsize{7}{11}\selectfont 81.31\%} 
& {\fontsize{7}{11}\selectfont 20.3} & {\fontsize{7}{11}\selectfont 71.72\%} \\
\hline
{\fontsize{8}{11}\selectfont BO} 
& {\fontsize{7}{11}\selectfont 9.55} & {\fontsize{7}{11}\selectfont 61.06\%} 
& {\fontsize{7}{11}\selectfont 3.86} & {\fontsize{7}{11}\selectfont 55.94\%}
& {\fontsize{7}{11}\selectfont 34.2} & {\fontsize{7}{11}\selectfont 56.36\%} 
& {\fontsize{7}{11}\selectfont 82.1} & {\fontsize{7}{11}\selectfont 63.8\%} 
& {\fontsize{7}{11}\selectfont 12.9} & {\fontsize{7}{11}\selectfont 82.52\%} 
& {\fontsize{7}{11}\selectfont 19.9} & {\fontsize{7}{11}\selectfont 72.64\%} \\
\hline
{\fontsize{8}{11}\selectfont PSO} 
& {\fontsize{7}{11}\selectfont 9.76} & {\fontsize{7}{11}\selectfont 60.2\%} 
& {\fontsize{7}{11}\selectfont 3.84} & {\fontsize{7}{11}\selectfont 56.22\%}
& {\fontsize{7}{11}\selectfont 37.2} & {\fontsize{7}{11}\selectfont 52.52\%} 
& {\fontsize{7}{11}\selectfont 80.5} & {\fontsize{7}{11}\selectfont 64.5\%} 
& {\fontsize{7}{11}\selectfont 13.9} & {\fontsize{7}{11}\selectfont 81.22\%} 
& {\fontsize{7}{11}\selectfont 20.2} & {\fontsize{7}{11}\selectfont 71.95\%} \\
\hline
{\fontsize{8}{11}\selectfont GS} 
& {\fontsize{7}{11}\selectfont 8.82} & {\fontsize{7}{11}\selectfont 64.01\%} 
& {\fontsize{7}{11}\selectfont 3.37} & {\fontsize{7}{11}\selectfont 61.53\%} 
& {\fontsize{7}{11}\selectfont 30.11} & {\fontsize{7}{11}\selectfont 61.6\%} 
& {\fontsize{7}{11}\selectfont 73.74} & {\fontsize{7}{11}\selectfont 67.5\%} 
& {\fontsize{7}{11}\selectfont 12.46} & {\fontsize{7}{11}\selectfont 83.13\%} 
& {\fontsize{7}{11}\selectfont 17.94} & {\fontsize{7}{11}\selectfont 75.03\%} \\
\hline
\multicolumn{13}{|c|}{The Proposed Hyper-parameter Optimization for ADNN}\\
\hline
{\fontsize{8}{11}\selectfont AGA} 
& {\fontsize{7}{11}\selectfont 9.63} & {\fontsize{7}{11}\selectfont 60.71\%} 
& {\fontsize{7}{11}\selectfont 4.26} & {\fontsize{7}{11}\selectfont 51.32\%} 
& {\fontsize{7}{11}\selectfont 36.7} & {\fontsize{7}{11}\selectfont 53.24\%} 
& {\fontsize{7}{11}\selectfont 83.6} & {\fontsize{7}{11}\selectfont 63.15\%} 
& {\fontsize{7}{11}\selectfont 13.4} & {\fontsize{7}{11}\selectfont 81.91\%} 
& {\fontsize{7}{11}\selectfont 20.6} & {\fontsize{7}{11}\selectfont 71.28\%} \\
\hline
{\fontsize{8}{11}\selectfont GA+KNN} 
& {\fontsize{7}{11}\selectfont 9.08} & {\fontsize{7}{11}\selectfont 62.97\%} 
& {\fontsize{7}{11}\selectfont 3.4} & {\fontsize{7}{11}\selectfont 61.17\%} 
& {\fontsize{7}{11}\selectfont 30.11} & {\fontsize{7}{11}\selectfont 61.6\%} 
& {\fontsize{7}{11}\selectfont 74.34} & {\fontsize{7}{11}\selectfont 67.23\%} 
& {\fontsize{7}{11}\selectfont 12.46} & {\fontsize{7}{11}\selectfont 83.13\%} 
& {\fontsize{7}{11}\selectfont 18.09} & {\fontsize{7}{11}\selectfont 74.78\%} \\
\hline
{\fontsize{8}{11}\selectfont Our method} 
& {\fontsize{7}{11}\selectfont 9.09} & {\fontsize{7}{11}\selectfont 62.91\%} 
& {\fontsize{7}{11}\selectfont 3.43} & {\fontsize{7}{11}\selectfont 60.89\%} 
& {\fontsize{7}{11}\selectfont 30.11} & {\fontsize{7}{11}\selectfont 61.6\%} 
& {\fontsize{7}{11}\selectfont 74.75} & {\fontsize{7}{11}\selectfont 67.05\%} 
& {\fontsize{7}{11}\selectfont 12.46} & {\fontsize{7}{11}\selectfont 83.13\%} 
& {\fontsize{7}{11}\selectfont 18.17} & {\fontsize{7}{11}\selectfont 74.67\%} \\

\hline 
\end{tabular}
\end{center}
\vspace{-2mm}
\end{table*}

\begin{figure*}
\setlength{\abovecaptionskip}{-0.05cm} 
	\centering
	\subfigure[]{
		\begin{minipage}[t]{0.31\linewidth}
			\centering
			\includegraphics[width=2.4in]{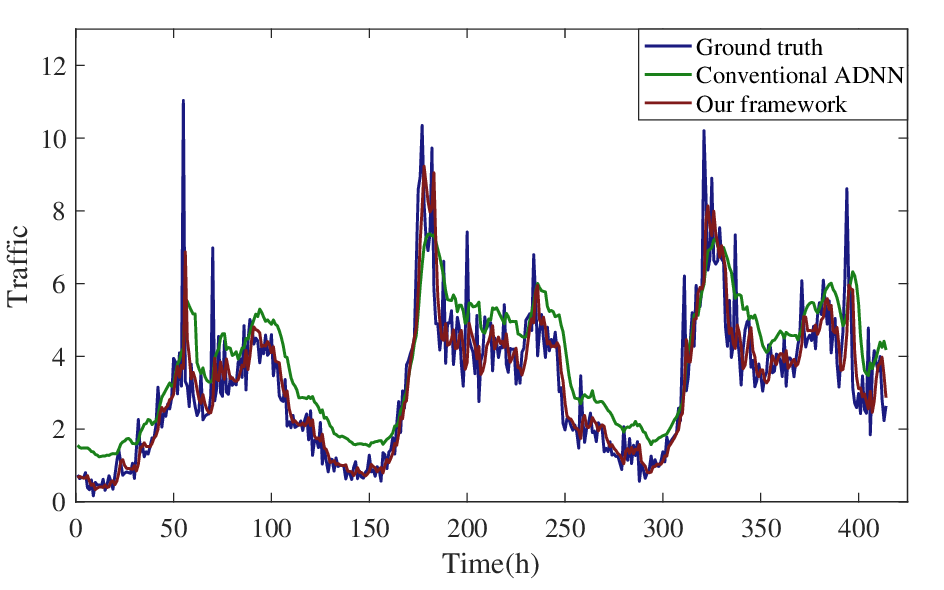}
		\end{minipage}
	}
	\subfigure[]{
		\begin{minipage}[t]{0.31\linewidth}
			\centering
			\includegraphics[width=2.4in]{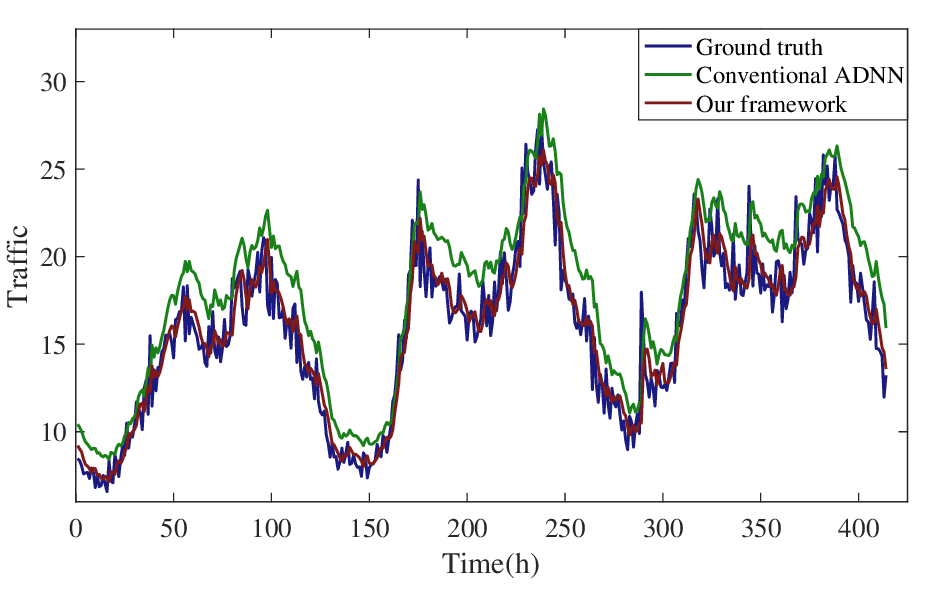}
		\end{minipage}
	}
	\subfigure[]{
		\begin{minipage}[t]{0.31\linewidth}
			\centering
			\includegraphics[width=2.4in]{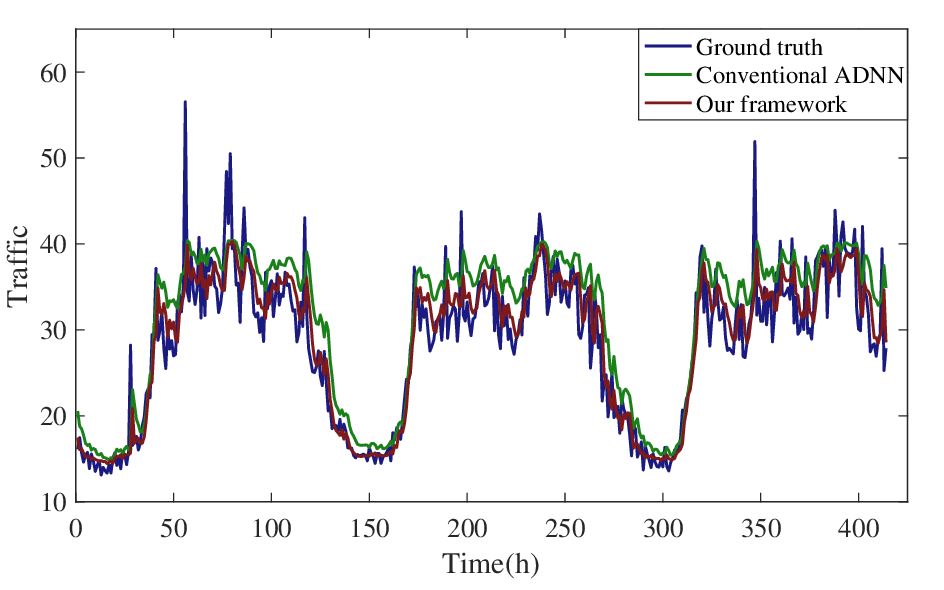}
		\end{minipage}
	}
	\centering
	\caption{Predicted traffic loads of the conventional ADNN and our framework as well as the ground true traffic loads generated in three real-world mobile cells: a) mobile cell 785; b) mobile cell 6708; c) mobile cell 9106.}
	\label{fig11}
\vspace{-2.5mm}
\end{figure*}

The proposed framework achieves high prediction accuracy and demonstrates stable performance across testing base-tasks. This can be attributed to two key factors.
First, the base-learner in the proposed framework is designed as the ADNN, which exhibits the potential for effectively capturing complex temporal correlations and nonlinear patterns in cell-level wireless network traffic loads.
Second, unlike conventional deep learning-based prediction models that rely on manual tuning or randomly selected hyper-parameters, the proposed framework employs a meta-learner for automated hyper-parameter optimization. By leveraging the meta-features of each base-task, the meta-learner constructs a hypothesis space containing appropriate functions that closely resemble the target function of the given task. 
Then, it efficiently determines the optimal hypothesis function based on the iterative search of AGA and rapid screening for offspring chromosomes, further improving both prediction accuracy and computational efficiency.

Compared to traditional hyper-parameter optimization methods including SS, GA, BO, and PSO, the proposed framework consistently delivers superior hyper-parameter selection strategies for testing base-learners while maintaining stable optimization performance. 
This advantage stems from the hyper-parameter optimization experience accumulated from meta-samples, which enhances both the KNN learning algorithm and AGA. Specifically, the KNN learning algorithm generates high-quality first-generation chromosomes, serving as strong initial search points for each hyper-parameter optimization problem. Meanwhile, it enables the subsequent AGA to explore the vast solution space efficiently and converge to a satisfactory solution. 
As shown in Table \ref{tab2}, both GA+KNN and GS fail to significantly improve the base-learners' post-training performance compared to our framework, and they also require considerably more computational time. 
This result highlights that the proposed framework efficiently generates highly appropriate solutions for each base-learner, closely approximating the theoretically optimal solution. It also indicates that the chromosome screening module effectively screens high-quality offspring chromosomes in each generation, even though the practical fitness values of those offspring chromosomes are not directly calculated. The computational time of these methods will be further compared in detail in Section V-E.

\vspace{0.2mm}
Figs. \ref{fig11} (a), (b), and (c) show the predicted traffic loads for the conventional ADNN, the proposed framework, and the ground truth traffic loads for three randomly selected testing mobile cells: 785, 6708, and 9106. It is evident that the proposed framework enables the base-learners to adapt to different base-tasks and accurately predict cell-level wireless network traffic loads by providing appropriate hyper-parameter selection strategies. In contrast, the conventional ADNNs with randomly selected hyper-parameters for cells 785 and 6708 exhibit significant prediction errors, despite being well trained. This highlights the crucial role of hyper-parameter optimization in NTP models and underscores the significance of this research.

\vspace{1mm}
\subsection{Computational time of the proposed framework}
\begin{figure}[!t]
\setlength{\abovecaptionskip}{-0.04cm} 
\centerline{\includegraphics[width= 3.1in]{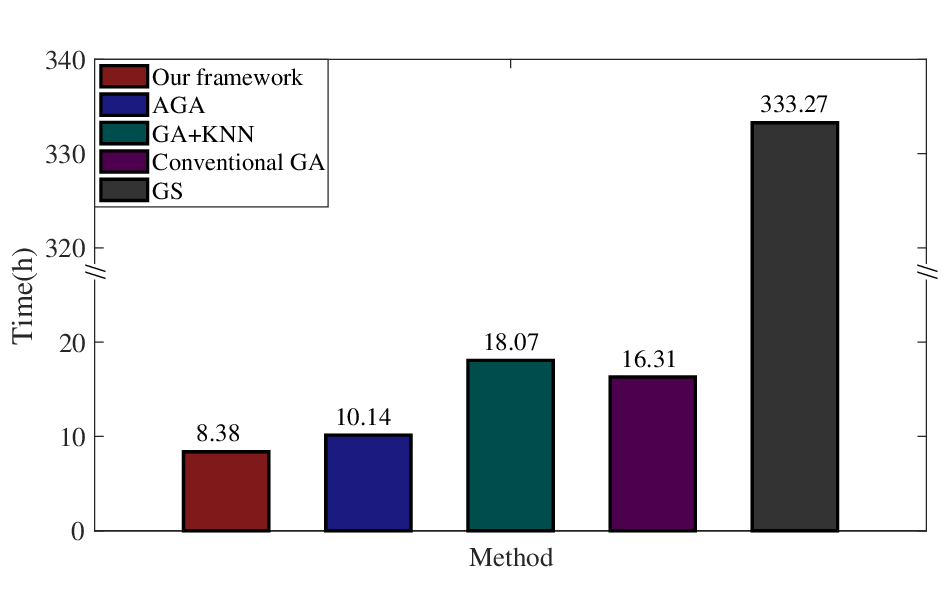}}
\caption{The average on-line computational time of the considered hyper-parameter optimization methods.}
\label{fig:cvgtimefor20cells}
\vspace{-4.5mm}
\end{figure}

Fig. \ref{fig:cvgtimefor20cells} illustrates the average online computational time for various hyper-parameter optimization methods, including the proposed framework, GA, AGA, GA+KNN, and GS, across the testing base-tasks.

From Fig. 12, it is evident that the GS incurs the highest computational time. This is because GS requires calculating the fitness value of each hyper-parameter selection strategy for the base-learner of every testing base-task, resulting in the base-learner being well-trained as many times as there are possible hyper-parameter strategies. Clearly, when the hyper-parameter optimization problem has a large solution space, GS may become impractical due to its significant computational complexity.

In contrast, the proposed framework and AGA benefit from a novel deep learning-assisted chromosome screening module, allowing them to quickly find out surviving offspring chromosomes without calculating all of the exact fitness values in each iteration (chromosome generation). As a result, the proposed framework and AGA exhibit much lower online computational complexities compared to GA and GA+KNN. Notably, the computational complexity advantage of the proposed framework and AGA over GA and GA+KNN further increases as the ratio between $W$ and $M$ grows.

An interesting observation from Fig. \ref{fig:cvgtimefor20cells} is that the proposed framework has lower computational complexity than AGA, even though both methods employ the intelligent chromosome screening module. This advantage arises because the KNN learning algorithm provides high-quality first-generation chromosomes, 
making it more likely that offspring chromosomes will survive in the early stages of \textbf{Algorithm 1}, thus reducing the need for extensive fitness value evaluations in subsequent iterations.

\begin{figure}[!t]
\setlength{\abovecaptionskip}{-0.04cm} 
\centerline{\includegraphics[width= 3.1in]{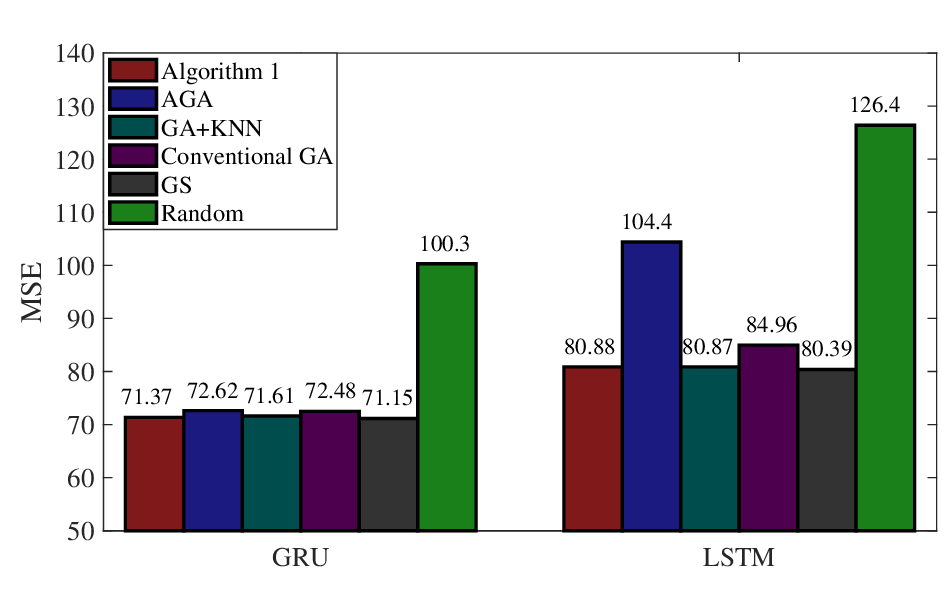}}
\caption{MSE performance achieved by the considered hyper-parameter optimization methods when base-learners adopt other deep learning algorithms rather than ADNN.}
\label{fig:GRU_LSTM_MSE}
\vspace{-4.5mm}
\end{figure}

Please note that the proposed framework requires a relatively long computational time to construct the set of meta-samples for its meta-learner. However, since these meta-samples are obtained offline and the meta-knowledge only needs to be prepared once, this additional offline computational complexity can be justified by the enhanced post-training prediction accuracy of the base-learners.

\vspace{-1mm}
\subsection{Robustness analyses of \textbf{Algorithm 1}}
The robustness of \textbf{Algorithm 1} is evaluated where base-learners adopt other deep learning-based models for cell-level wireless NTP.

\begin{figure}[!t]
\setlength{\abovecaptionskip}{-0.04cm} 
\centerline{\includegraphics[width= 3.1in]{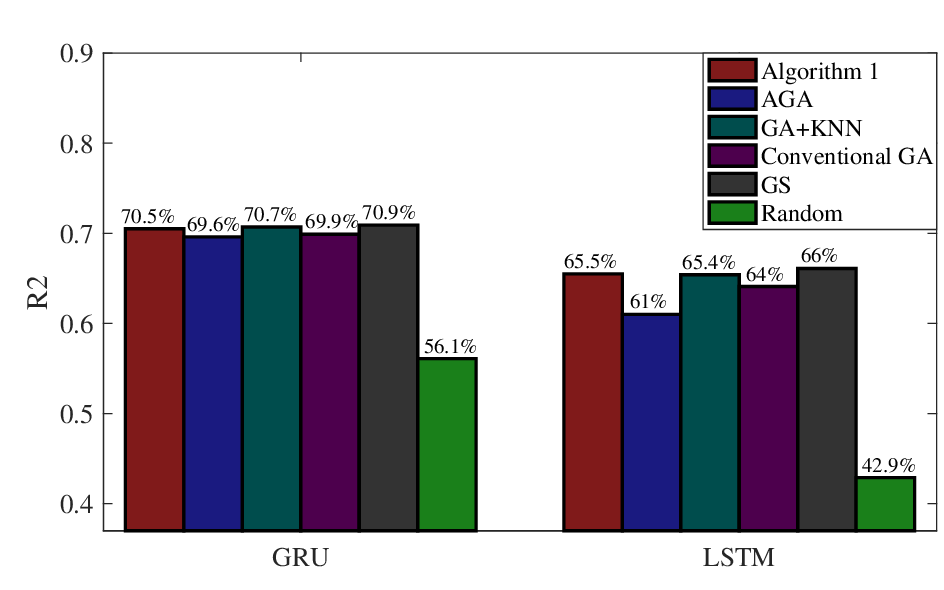}}
\caption{R2 performance achieved by the considered hyper-parameter optimization methods when base-learners adopt other deep learning algorithms rather than ADNN.}
\label{fig:GRU_LSTM_R2}
\vspace{-4mm}
\end{figure}

\begin{figure}[!t]
\setlength{\abovecaptionskip}{-0.04cm} 
\centerline{\includegraphics[width= 3.1in]{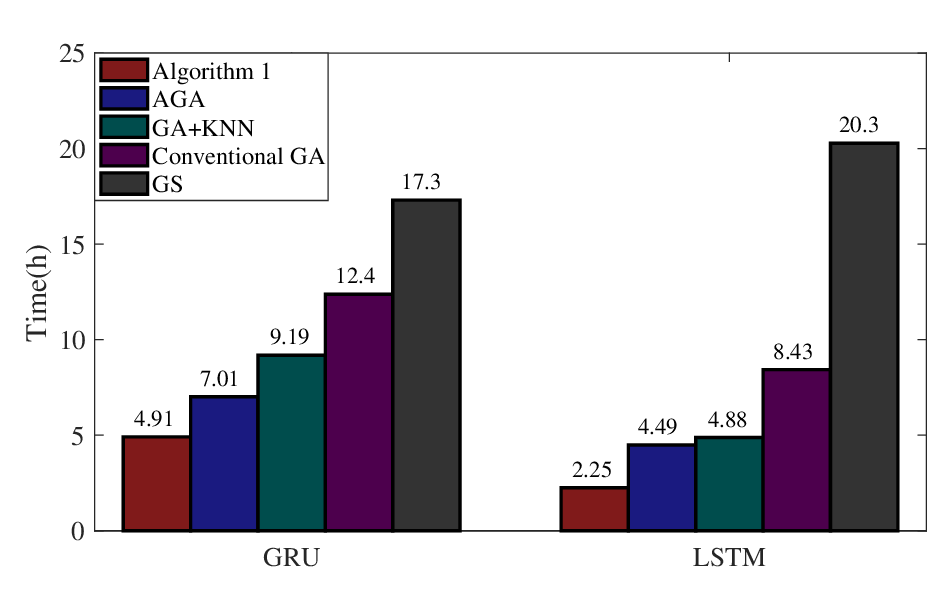}}
\caption{The average on-line computational time of the considered hyper-parameter optimization methods when base-learners adopt other deep learning algorithms rather than ADNN.}
\label{fig:GRU_LSTM_Time}
\vspace{-2mm}
\end{figure}

Specifically, the cases are examined where base-learners are implemented as either LSTM or GRU networks, with the corresponding hyper-parameter selection ranges detailed in Table \ref{tab1}. Figs. \ref{fig:GRU_LSTM_MSE} and \ref{fig:GRU_LSTM_R2} show the extent to which the proposed framework improves the post-training prediction accuracy of base-learners, evaluated in terms of the average MSE and R2 coefficient across the testing base-tasks. Additionally, Fig. \ref{fig:GRU_LSTM_Time} compares the online computational time of \textbf{Algorithm 1} against other hyper-parameter optimization methods.
For each deep learning model, an identical set of meta-samples is constructed for \textbf{Algorithm 1}, where each chromosome represents a feasible hyper-parameter selection strategy for the respective base-learner. From Figs. \ref{fig:GRU_LSTM_MSE}, \ref{fig:GRU_LSTM_R2}, and \ref{fig:GRU_LSTM_Time}, it is evident that, compared to randomly chosen hyper-parameters, \textbf{Algorithm 1} consistently reduces the average MSE, increases the average R2 coefficient of base-learners, and maintains an acceptable online computational cost, regardless of the deep learning model employed by base-learner. Moreover, in comparison to the GS, \textbf{Algorithm 1} consistently determines hyper-parameter selection strategies that yield post-training prediction accuracies closely matching those obtained with theoretically optimal hyper-parameters, while maintaining high computational efficiency. The results presented in Figs. \ref{fig:GRU_LSTM_MSE}, \ref{fig:GRU_LSTM_R2}, and \ref{fig:GRU_LSTM_Time} collectively demonstrate the robustness of \textbf{Algorithm 1} across different deep learning models employed by the base-learners.

\vspace{-2mm}
\section{Conclusion}  
In this paper, a novel meta-learning based hyper-parameter optimization framework is proposed for wireless NTP models. 
Within the proposed framework, an ADNN is designed as the base-learner to handle each wireless NTP task, referred to as a base-task. 
The hyper-parameter optimization process for each base-learner is formulated as a meta-task, characterized by the meta-features of its corresponding base-task, and is addressed through a novel meta-learner. 
Based on observations from real-world traffic data, where base-tasks with similar meta-features tend to prefer similar hyper-parameters for their base-learners, the meta-learner employs a KNN learning algorithm to generate a set of high-quality candidate hyper-parameter selection strategies for a new base-learner by leveraging meta-knowledge. 
Subsequently, an AGA equipped with a deep learning-assisted intelligent chromosome screening module is utilized to determine the optimal hyper-parameters. 
The proposed framework is evaluated on real-world wireless NTP tasks. 
Extensive experiments demonstrate that it significantly improves the post-training prediction accuracy of base-learners by providing near-optimal hyper-parameter selection strategies. Furthermore, results indicate that the proposed framework exhibits strong robustness across different deep learning models adopted by base-learners. 
Source codes of this work can be found at \textit{https://github.com/LZLZLLL/HO-CWNTP/}.

\appendices
\renewcommand{\theequation}{\thesection--\arabic{equation}}

\ifCLASSOPTIONcaptionsoff
  \newpage
\fi

\end{document}